\documentclass[conference]{IEEEtran}

\usepackage{cite}
\usepackage{amsmath,amssymb,amsfonts}
\usepackage{algorithmic}
\usepackage{graphicx}
\usepackage{textcomp}
\usepackage{xcolor}

\usepackage{url}
\usepackage{hyperref}
\usepackage{multirow}
\usepackage{listings}
\usepackage{booktabs}
\usepackage[most]{tcolorbox}
\usepackage{tagging}
\usepackage{ipdps26repro}
\usepackage{annotate-equations}

\def\BibTeX{{\rm B\kern-.05em{\sc i\kern-.025em b}\kern-.08em
    T\kern-.1667em\lower.7ex\hbox{E}\kern-.125emX}}

\newcommand{\tweakedsim}{\raise.17ex\hbox{$\scriptstyle\mathtt{\sim}$}}

\definecolor{dkgreen}{RGB}{0,64,0}
\definecolor{ltgray}{RGB}{245,245,245}
\definecolor{mauve}{RGB}{139,0,139}

\definecolor{color_1}{HTML}{009E73} 
\definecolor{color_2}{HTML}{D55E00} 
\definecolor{color_3}{HTML}{0072B2} 
\definecolor{color_4}{HTML}{000000} 
\definecolor{color_5}{HTML}{E69F00} 
\definecolor{color_6}{HTML}{CC79A7} 

\definecolor{belizehole}{HTML}{2980b9}
\newtcolorbox{RQcallout}[1][]{%
    colback=black!5,
    colframe=black!5,
    notitle,
    sharp corners,
    borderline west={2pt}{0pt}{belizehole!80!black},
    enhanced,
    breakable,
  }

\newcommand{\RQtwo}{Are there differences in how GPU jobs utilize FP16, FP32, FP64, and Tensor pipes?}
\newcommand{\RQthree}{Among the jobs that request 80 GB GPUs, how much HBM do they actually use?}

\newcommand{\RQfive}{What share of jobs are classified as compute-bound versus memory-bound under the roofline criterion, and how do these job types differ in total energy consumption?}
\newcommand{\RQsix}{Do jobs that utilize multiple GPUs use them evenly, and how does spatial imbalance vary with overall GPU utilization, FP pipes used, and job type (compute- versus memory-bound)?}
\newcommand{\RQseven}{
Are GPUs used consistently over time 
within a single job, and how does temporal imbalance vary with overall GPU utilization, FP pipes used, and job type (compute- versus memory-bound)?
}
\newcommand{\RQeight}{How do compute and memory activity counters relate to overall GPU utilization across compute-bound and memory-bound jobs, and which counter pairs exhibit the strongest correlations?}

\newcommand{\RQoneAnswer}{Overall, we find that small node-count jobs dominate Perlmutter's workload. Median runtimes are shorter for single-node jobs compared to multi-node jobs. Typical durations for multi-node jobs are around 1–3 hours. Overall, GPU utilization is highest at intermediate job sizes (33–512 GPUs) and lower at the largest sizes ($\ge$512 GPUs).}
\newcommand{\RQtwoAnswer}{The FP64-only group is the largest (44\% of jobs), which indicates that many workloads utilize FP64 pipes on Perlmutter. In addition, jobs that utilize tensor pipes alongside FP64 or FP32 tend to sustain the highest GPU utilization (50\% and 46\%, respectively), which suggests that utilizing tensor cores tends to improve GPU utilization.}
\newcommand{\RQthreeAnswer}{Among jobs that explicitly request 80 GB GPUs, 55\% peak at or below 50\% of HBM capacity and should fit on a 40 GB GPU. Meanwhile, 20\% of jobs exhibit good capacity use with HBM usage of at least 80\%.}

\newcommand{\RQfiveAnswer}{The jobs on Perlmutter are overwhelmingly memory-bound (\tweakedsim 80\% of all jobs). Memory-bound jobs typically consume more energy compared to compute-bound jobs.}

\newcommand{\RQsevenAnswer}{
Greater mean GPU\_UTIL correlates with lower temporal imbalance. Low-utilization jobs often exhibit high temporal imbalance, whereas high-utilization jobs maintain consistently high GPU\_UTIL (low temporal imbalance). FP64-only jobs have higher temporal imbalance. Compute-bound jobs exhibit more temporal imbalance than memory-bound jobs. 
}
\newcommand{\RQeightAnswer}{In compute-bound jobs, GPU utilization is most strongly associated with compute activity, while DRAM activity contributes weakly. In memory-bound jobs, both compute and DRAM activity are positively associated with utilization. Across all jobs, overall GPU utilization aligns most closely with SM activity, power usage, and GPU temperature.}

\makeatletter
\newcommand{\linebreakand}{%
  \end{@IEEEauthorhalign}
  \hfill\mbox{}\par
  \mbox{}\hfill\begin{@IEEEauthorhalign}
}
\makeatother

\begin{document}

\title{Characterizing Production GPU Workloads using System-wide Telemetry Data}

\author{
  \IEEEauthorblockN{Onur Cankur}
  \IEEEauthorblockA{\textit{Department of Computer Science}\\
    \textit{University of Maryland}\\
    College Park, Maryland 20742 USA\\
    ocankur@umd.edu
  }
  \and
  \IEEEauthorblockN{Brian Austin}
  \IEEEauthorblockA{\textit{NERSC}\\
    \textit{Lawrence Berkeley National Laboratory}\\
    Berkeley, California 94720 USA\\
    baustin@lbl.gov
  }
  \linebreakand
  \IEEEauthorblockN{Dhruva Kulkarni}
  \IEEEauthorblockA{\textit{NERSC}\\
    \textit{Lawrence Berkeley National Laboratory}\\
    Berkeley, California 94720 USA\\
    dkulkarni@lbl.gov
  }
  \and
  \IEEEauthorblockN{Abhinav Bhatele}
  \IEEEauthorblockA{\textit{Department of Computer Science}\\
    \textit{University of Maryland}\\
    College Park, Maryland 20742 USA\\
    bhatele@cs.umd.edu
  }
}

\maketitle

\begin{abstract}
  GPGPU-accelerated clusters and supercomputers are central to modern high-performance computing (HPC). Over the past decade, these systems continue to expand, and GPUs now expose a wide range of hardware counters that provide detailed views of performance and resource usage. Despite the potential of these counters, few studies have evaluated the insights they offer about real workloads at scale. In this work, we address this gap by analyzing previously underexplored GPU hardware counters collected via Lightweight Distributed Metric Service on Perlmutter, a leadership-class supercomputer. We quantify uneven work distribution across GPUs within a job and the steadiness of GPU activity over time, and we classify jobs as compute- or memory-bound using a roofline-based criterion. We then use these metrics to interpret job behavior in terms of practical workload characteristics to provide interpretable, job-level insights. Our findings can inform workload optimization and future HPC system design. For example, 81\% of jobs are memory-bound, and memory-bound jobs tend to consume more energy than compute-bound jobs at comparable GPU-hours. Among jobs requesting 80 GB GPUs, 55\% peak at 50\% HBM capacity or less.

\end{abstract}

\begin{IEEEkeywords}
  workload characterization, GPU telemetry, resource utilization
\end{IEEEkeywords}

\section{Introduction}
\label{sec:intro}
General-purpose Graphics Processing Units (GPGPUs) have become pervasive in
compute nodes on high-performance computing (HPC) systems. Understanding GPU resource usage has the potential to
inform application optimization, facility operation, and future architecture design. This is enabled by system-wide monitoring datasets collected over extended periods across all compute nodes, which capture GPU floating-point (FP) pipe activity, HBM usage, interconnect traffic, and energy consumption together with job metadata such as start and end times and GPU counts.

Although modern GPUs expose a rich set of hardware counters, operators and users rarely exploit this breadth. Not leveraging this data creates performance blind spots: users struggle to diagnose underperformance and to choose effective optimizations for their jobs, and system operators lack the observability needed for informed placement and capacity planning and detecting inefficiencies at scale. To address this underuse, we conduct a telemetry-based study on Perlmutter. The aim of this study is to translate system-level hardware counter telemetry into interpretable, job-level insights that inform both users and operators.

System-wide monitoring data requires a rigorous curation and analysis workflow to produce valid job-level insights. Reliable analysis begins with precise alignment of each GPU sample to its job metadata. Because hardware counters differ in semantics (some report fractional activity, others provide instantaneous readings, and others accumulate totals), the workflow enforces correct interpretations, handles missing values and outliers, and applies appropriate aggregations to produce job-level summaries with robust methods that support statistically defensible comparisons.

While some prior work analyzes GPU monitoring data~\cite{chu2024generic, li2023analyzing, li2022ai}, most studies consider only a limited set of hardware counters. We address this gap with a comprehensive study of GPU-specific counters collected by the Lightweight Distributed Metric Service (LDMS) on Perlmutter, a flagship open-science supercomputer at the National Energy Research Scientific Computing Center (NERSC). We analyze one month of data sampled every 10 seconds from March 1 to April 1, 2025 across all compute nodes. The dataset includes NVIDIA Data Center GPU Manager (DCGM) counters covering SM activity, FP32/FP64/tensor pipe activity, frame-buffer (HBM) usage, NVLink and PCIe traffic, energy, and temperature. A one-month window is commonly used as representative of system behavior~\cite{michelogiannakis2022case}.

\begin{table*}[t]
    \centering
    \caption{DCGM hardware counters used in this study and their descriptions. All the counter names below have a DCGM\_FI\_ prefix, which has been omitted from the table for brevity.}
    \label{table:dataset_info}
    {\footnotesize
        \begin{tabular}{llp{9.4cm}} \toprule
            \textbf{Counter Name (DCGM\_FI\_X)}                   & \textbf{Short Name} & \textbf{Description}                                                                           \\  \midrule
            DEV\_GPU\_UTIL                  & GPU\_UTIL           & Fraction of time during which at least one kernel was executing on the GPU.                                          \\
            PROF\_SM\_ACTIVE                & SM\_ACTV            & Fraction of time at least one warp was active, averaged over all multiprocessors.                                    \\
            PROF\_PIPE\_FP16\_ACTIVE        & FP16\_ACTV          & Fraction of cycles the FP16 (half-precision) pipe was active.                                                        \\
            PROF\_PIPE\_FP32\_ACTIVE        & FP32\_ACTV          & Fraction of cycles the FP32 (single-precision and integer) pipe was active.                                          \\
            PROF\_PIPE\_FP64\_ACTIVE        & FP64\_ACTV          & Fraction of cycles the FP64 (double-precision) pipe was active.                                                      \\
            PROF\_PIPE\_TENSOR\_ACTIVE      & TNSR\_ACTV          & Fraction of cycles the tensor pipe was active.                                                                       \\
            PROF\_DRAM\_ACTIVE              & DRAM\_ACTV          & Fraction of cycles where data was sent to or received from device memory.                                            \\
            DEV\_FB\_USED                   & HBM\_USED           & Absolute amount high-bandwidth memory (HBM) capacity used (MB).                                                      \\
            PROF\_NVLINK\_\{TX/RX\}\_BYTES  & NVLINK\_\{TX/RX\}   & Rate of data transmitted/received over NVLink, not including protocol headers, in bytes per second.                  \\
            PROF\_PCIE\_\{TX/RX\}\_BYTES    & PCIE\_\{TX/RX\}     & Rate of data transmitted/received over PCIe, including both protocol headers and data payloads, in bytes per second. \\
            DEV\_TOTAL\_ENERGY\_CONSUMPTION & TOTAL\_ENG          & Total energy consumption for the GPU in mJ since the driver was last reloaded.                                       \\
            DEV\_GPU\_TEMP                  & GPU\_TEMP           & Current temperature readings for the device, in degrees Celsius.                                                     \\ \bottomrule
        \end{tabular}
    }
\end{table*}

We present a detailed analysis of how GPU workloads use resources across GPUs and over time. First, we quantify how evenly work within a job is distributed across its assigned GPUs. When one GPU carries much more load than its peers while others idle, the job wastes capacity. Our measures capture this unevenness to assess the efficiency of multi-GPU jobs. Second, we quantify how steadily each GPU is exercised during a job's runtime.
Large swings indicate inconsistent use, whereas flatter profiles indicate sustained use.

We classify jobs as compute-bound or memory-bound using a roofline-based criterion to determine whether performance is limited by compute or memory, and we relate the evenness and steadiness of use to these job types. We further examine usage of FP pipes (FP16, FP32, FP64, tensor), HBM memory, and the NVLink and PCIe interconnects, and we analyze correlations between hardware counters and GPU utilization to identify factors associated with higher GPU use. We describe the dataset and preprocessing, define the metrics used throughout, and present results that address our research questions. We conclude with implications and future directions.

Specifically, this work makes the following contributions:
\begin{itemize}
    \item We conduct a system-wide analysis of 75,703 jobs on Perlmutter over one month, using DCGM counters that are largely underexplored in prior work, with a robust job attribution pipeline. This enables statistically grounded insights into diverse aspects of GPU usage.
    \item We label jobs as compute-bound or memory-bound using a roofline criterion and observe that, at comparable GPU-hours, memory-bound jobs tend to consume more energy than compute-bound jobs.
    \item We introduce a time-windowed spatial imbalance metric that captures uneven resource usage across GPUs of a job, which enables minute-scale detection of uneven load and more reliable job-level summaries.
    \item We classify jobs by FP pipeline activity and quantify how GPU utilization and spatial and temporal imbalance vary across classes. We observe that tensor activity is typically associated with higher utilization.
    \item Using a variety of system-wide DCGM counters, we quantify job-level relationships and expose which counters co-vary with  GPU utilization and how these associations differ between compute and memory-bound jobs.
\end{itemize}

\section{Monitoring data used in this study}
\label{sec:data}
Perlmutter at the National Energy Research Scientific Computing Center (NERSC) is a heterogeneous system with both CPU-only and GPU-accelerated nodes. It comprises 3,072 CPU-only nodes and 1,792 GPU nodes. This study focuses exclusively on jobs submitted to the GPU partition. Each GPU node contains one AMD EPYC 7763 CPU, four NVIDIA A100 GPUs and 256 GB of DDR4 system memory. Among the GPU nodes, 1,536 use A100s with 40 GB of HBM per GPU and 256 use A100s with 80 GB of HBM per GPU. The four GPUs in each node are interconnected via third-generation NVLink (four links between each pair, 25 GB/s per direction per link), and the CPUs connect to both the GPUs and the network interface cards (NICs) over PCIe 4.0.

\subsection{Data sources}

Perlmutter uses LDMS~\cite{ldms} to collect system-wide monitoring data from compute nodes. LDMS provides plugins for different system components and continuously records hardware-counter measurements of CPU and GPU activity, memory utilization, and I/O. In this study, we retrieve performance-counter measurements from the Data Center GPU Manager (DCGM)~\cite{dcgm} plugin. We also retrieve Slurm~\cite{slurm_documentation} job scheduling data, which include job-specific metadata such as start and end times and allocated node lists.

We retrieve a month of data spanning March~1 to April~1, 2025. The DCGM plugin samples counters periodically on running jobs. The system admins configure the DCGM plugin to sample counters every 10 seconds on Perlmutter, and users cannot change this rate. Our dataset includes information about 75,703 jobs after the cleaning and preprocessing steps described in Section~\ref{subsec:preprocessing}.
We use only DCGM counters that Perlmutter staff validate for operational monitoring, since hardware counters can be noisy~\cite{ritter2022conquering}.

Table~\ref{table:dataset_info} lists the GPU counters used in this study. GPU\_UTIL is the fraction of time at least one kernel was executing on the device. SM\_ACTV is the fraction of time at least one warp was active (including stalled warps waiting on memory). FP16\_ACTV, FP32\_ACTV, FP64\_ACTV, and TNSR\_ACTV are activity fractions for the corresponding pipes; higher values indicate greater activity of that pipeline. DRAM\_ACTV is the fraction of cycles issuing device-memory traffic. NVLINK\_{TX,RX} and PCIE\_{TX,RX} report byte rates per second over NVLink (payload only) and PCIe (headers plus payload), respectively. HBM\_USED is the absolute HBM footprint (MB). TOTAL\_ENG is the cumulative GPU energy (mJ) since the driver was last reloaded. GPU\_TEMP is the current device temperature (Celsius).

\subsection{Data retrieval and integration}

We retrieve LDMS data from Perlmutter using the Prometheus API and
Slurm job data with the sacct command, and we store both in Parquet files.
LDMS data provides hardware counter values per GPU, including timestamps,
node IDs, GPU IDs, and counter values, but lacks job IDs. In contrast,
Slurm data contains job-level metadata without hardware counters.

To enable per-job analysis, we merge the datasets by aligning LDMS samples with Slurm data.
We match LDMS timestamps to jobs running on the same node using their start and end times from Slurm data.
For each job, we assign LDMS samples within its time range to the corresponding job ID and step.
This process links each LDMS sample to a specific job and step, which enables us to associate GPU-level hardware-counter values with specific jobs.

\subsection{Data filtering and preprocessing}
\label{subsec:preprocessing}
We apply several filtering and preprocessing steps to ensure reliable and
accurate analyses of the merged LDMS and Slurm datasets. First, we remove the
LDMS entries not associated with any jobs during the merging process. We exclude
the jobs running on the login nodes since they do not represent meaningful
workloads. We also remove jobs that do not use the GPU partition.

Additionally, we exclude the jobs submitted by staff accounts since these jobs
often involve maintenance, testing, or debugging rather than production
workloads. We remove the jobs with durations
shorter than three minutes, as they typically lack sufficient data for
meaningful analysis.
We also apply some counter-specific filters to address data inconsistencies.
For instance, we remove samples where counter values exceed their
physical limits (e.g., GPU\_UTIL \textgreater 100\%). These samples are extremely rare (less than 0.0001\% of all records). We suspect they are due to counter overflow, and we therefore remove them. Finally, we exclude jobs with a job-level mean GPU\_UTIL value below 1\%.

\section{Data analytics methodology}
\label{sec:method}
Our goal is to convert raw, per-sample GPU telemetry into interpretable job-level insights. We (i) classify job type (compute- versus memory-bound) using roofline-based sample labels aggregated to the job level, (ii) measure spatial imbalance to quantify how evenly work is distributed across allocated GPUs, and (iii) measure temporal imbalance to quantify how steadily a job sustains activity over time. We then relate these metrics to job-level counter means using heatmaps and Spearman correlations.

Together, these views support user diagnosis (e.g., properly sizing allocations) and operator decisions (e.g., placement under HBM headroom, expectations for power/thermals). The subsections that follow formalize each metric, define aggregation and normalization rules, and detail how we compute job-level summaries and correlations from the counters in Table~\ref{table:dataset_info}.

\subsection{Classifying job types}
\label{subsec:method_roofline}
The roofline model identifies whether compute or memory activity limits a kernel. It plots each kernel with arithmetic intensity on the x-axis and achieved flop/s on the y-axis. Arithmetic intensity is FLOPs per byte moved from HBM. Achieved rate is the measured flop/s during execution. Two limit lines appear on the same chart: a flat line at peak flop/s and a rising line equal to peak HBM bandwidth times arithmetic intensity. Their intersection is the ridge. Points left of the ridge are memory-bound, and points at or right are compute-bound. At any x-value, the lower limit sets the attainable flop/s.

We perform a roofline-based analysis, building on~\cite{austin2024system}, to label each telemetry sample as compute-bound or memory-bound and then aggregate these labels at the job level. The method uses the floating-point (FP) pipeline activity fractions (FP16\_ACTV, FP32\_ACTV, FP64\_ACTV, TNSR\_ACTV) and the memory activity fraction (DRAM\_ACTV) listed in Table~\ref{table:dataset_info}, together with peak device capabilities. We take peak flop/s and peak HBM bandwidth from the device specifications. Because the HBM peak depends on GPU capacity (40~GB or 80~GB), we infer the capacity from node identifiers.

For each sample, achieved flop/s and HBM bandwidth are estimated by scaling the architectural peaks with the observed activity fractions. For example, the achieved FP64 flop/s and HBM bandwidth are estimated as follows:

\begin{equation}
  \mathit{F64} = \mathrm{FP64\_ACTV} \times \mathrm{peak~FP64~flop/s}
\end{equation}
\begin{equation}
  \mathit{B} = \mathrm{DRAM\_ACTV} \times \mathrm{peak~HBM~BW}
\end{equation}

The per-sample arithmetic intensity is then
\vspace{0.2in}
\begin{equation}
  \eqnmarkbox[color_1]{ai_out}{\mathit{AI}_{\mathrm{fp64}}} = \frac{\eqnmarkbox[color_2]{ai_num}{\mathit{F64}}}{\eqnmarkbox[color_3]{ai_den}{\mathit{B}}}
\end{equation}
\annotate[yshift=.6em]{above,left}{ai_out}{FP64 arithmetic intensity}
\annotate[yshift=.6em]{right,right}{ai_num}{achieved FP64 flop/s}
\annotate[yshift=-.1em]{below,right}{ai_den}{achieved HBM bandwidth}
\vspace{0.2in}

We compare $\mathit{AI}_{\mathrm{fp64}}$ to the device balance point (ridge) and classify a sample as compute-bound if $\mathit{AI}_{\mathrm{fp64}}$ exceeds the balance point, and as memory-bound otherwise. The balance point is defined as the ratio of the device’s peak compute throughput (flop/s) for the FP pipeline of interest to the device’s peak memory bandwidth (bytes/s).
Finally, we categorize each job by the fractions of its compute-bound and memory-bound samples.

\subsection{Quantifying the spatial imbalance of jobs}

Spatial behavior refers to how a job's workload is distributed across
its assigned GPUs. We extend a prior spatial imbalance
metric \cite{li2023analyzing, peng2021holistic} with a time-windowed
approach that captures usage throughout the job rather than over its
full runtime at once. This time-windowed spatial imbalance metric is
useful because it quantifies how evenly a job's GPU usage is distributed
over time.

Let $\mathit{TC}(g,w) = \sum_{t=1}^{t_w} C_{g,t}$ be the sum of hardware
counter values $C_{g,t}$ over all timestamps in time window $w$ for GPU $g$.
We define the spatial imbalance of job $j$ in $w$ as

\vspace{0.3in}
\begin{equation}
  \label{eq:si}
  \eqnmarkbox[color_1]{si_out}{\mathit{SI}(j,w)} =
  1 - \frac{
    \eqnmarkbox[color_2]{si_num}{\sum_{g=1}^{g_j} \mathit{TC}(g,w)}
  }{
    \eqnmarkbox[color_3]{si_den}{g_j \times \max\limits_{1 \leq g \leq g_j} \mathit{TC}(g,w)}
  }
\end{equation}
\annotate[yshift=.6em,xshift=.8em]{above,left}{si_out}{spatial imbalance\\in window $w$}
\annotate[yshift=.6em]{above,right}{si_num}{total activity\\in window $w$}
\annotate[yshift=-.3em]{below,left}{si_den}{max possible total\\given busiest GPU}
\vspace{0.3in}

The numerator in Equation~\ref{eq:si} sums the per-GPU hardware counter values within
time window $w$ across all GPUs allocated to the job where $g_j$ is the number of GPUs
allocated to $j$.
The denominator represents the maximum possible
counter value if all GPUs matched the highest observed value in that
window.  An $\mathit{SI}$ value near 0 indicates minimal imbalance; a value near 1
indicates significant imbalance.  We define the spatial imbalance
metric for a job over its entire runtime as the mean spatial
imbalance across all time windows, where $w_j$ is the total number of time
windows:
\vspace{0.2in}
\begin{equation}
  \eqnmarkbox[color_1]{siJ_out}{\mathit{SI}(j)} =
  \frac{
    \eqnmarkbox[color_2]{siJ_num}{\sum_{w=1}^{w_j} \mathit{SI}(j,w)}
  }{
    \eqnmarkbox[color_3]{siJ_den}{w_j}
  }
\end{equation}
\annotate[yshift=.6em]{above,left}{siJ_out}{job-level\\ spatial imbalance}
\annotate[yshift=.6em]{above,right}{siJ_num}{sum SI over time windows}
\annotate[yshift=-.3em]{below,right}{siJ_den}{number of windows}
\vspace{0.2in}

A longer time window smooths out short bursts by averaging them
over a longer period. In contrast, a shorter window retains these quick
changes and makes them more visible in the metric. Window length therefore sets the time scale of the imbalance we capture. We keep the window length fixed when comparing jobs so the metric is consistent across workloads.

\subsection{Quantifying the temporal imbalance of jobs}
Temporal imbalance quantifies the variation in hardware counter values over a job's
runtime.
We adopt the definition
from~\cite{peng2021holistic}. Here, $ C_{g,t} $ is the hardware counter value for GPU $g$ at time $t$. The numerator
sums the counter values over time. The denominator represents the maximum
possible value and is calculated by
multiplying the job duration by the peak observed counter value for that GPU.
Subtracting from 1 highlights the imbalance, where values near zero
indicate stable behavior and higher values reflect greater
fluctuations.

\vspace*{0.3in}
\begin{equation}
  \label{eq:ti}
  \eqnmarkbox[color_1]{ti_out}{\mathit{TI}(j,g)} =
  1 - \frac{
    \eqnmarkbox[color_2]{ti_num}{\sum_{t=1}^{t_j} C_{g,t}}
  }{
    \eqnmarkbox[color_3]{ti_den}{t_j \times \max_{1 \le t \le t_j} C_{g,t}}
  }
\end{equation}
\annotate[yshift=.6em]{above,left}{ti_out}{temporal imb.\\for GPU $g$}
\annotate[yshift=.6em]{above,right}{ti_num}{total activity\\over time}
\annotate[yshift=-.3em]{below,right}{ti_den}{max possible total\\given peak activity}
\vspace{0.3in}

Equation~\ref{eq:ti} defines the temporal imbalance metric for a single GPU.
We define the temporal imbalance of a job as the maximum imbalance across
all allocated GPUs, where $g_j$ is the number of GPUs assigned to the job:

\vspace{0.2in}
\begin{equation}
  \eqnmarkbox[color_1]{tiJ_out}{\mathit{TI}(j)} =
  \eqnmarkbox[color_2]{tiJ_max}{\max_{1 \le g \le g_j}\mathit{TI}(j,g)}
\end{equation}
\annotate[yshift=.8em]{above}{tiJ_out}{job-level temporal imbalance}
\annotate[yshift=-.3em]{below,left}{tiJ_max}{largest per-GPU temporal imbalance}
\vspace{0.1in}

\begin{figure*}[t]
  \centering
  \includegraphics[height=1.68in]{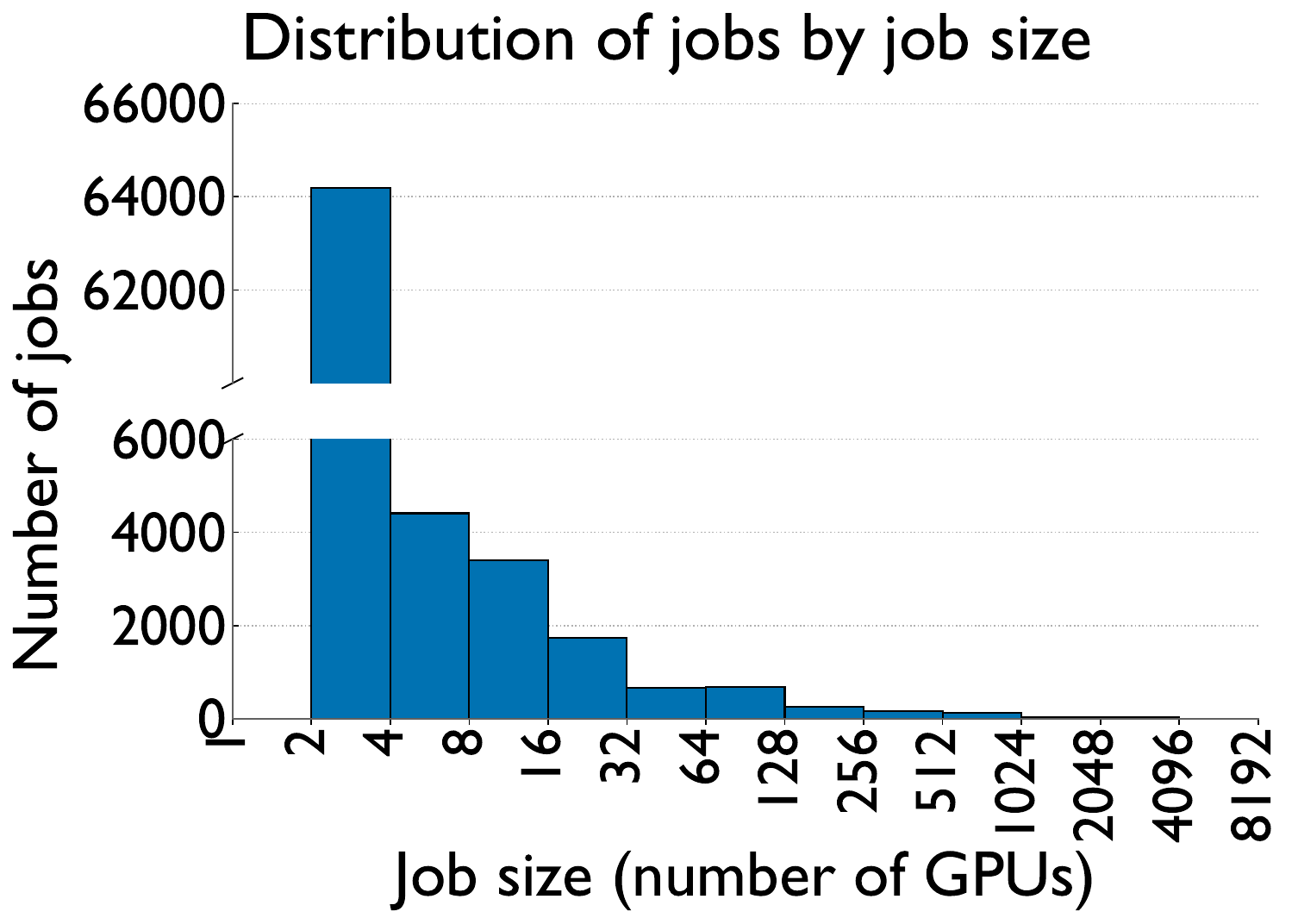}
  \includegraphics[height=1.68in]{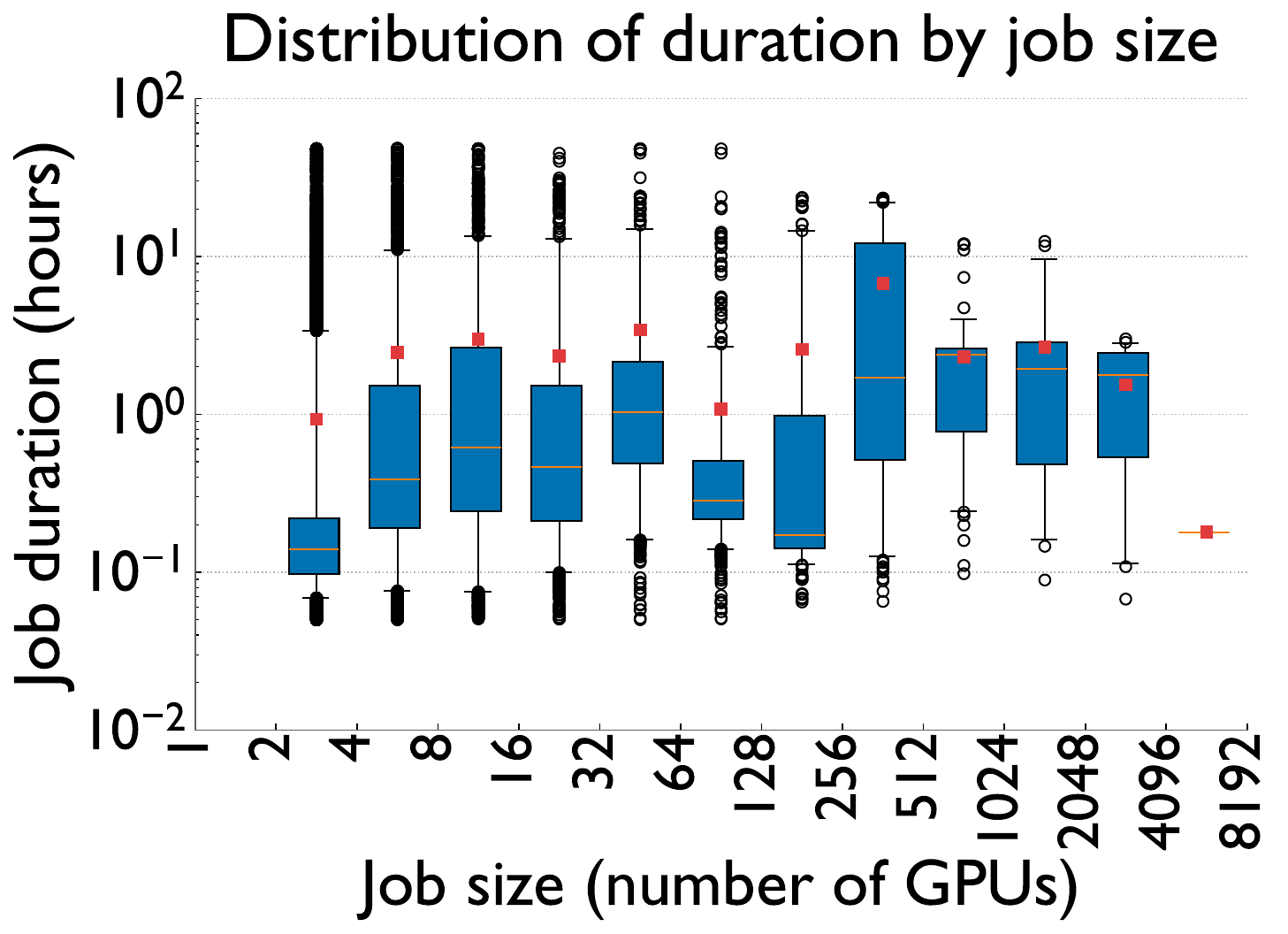}
  \includegraphics[height=1.68in]{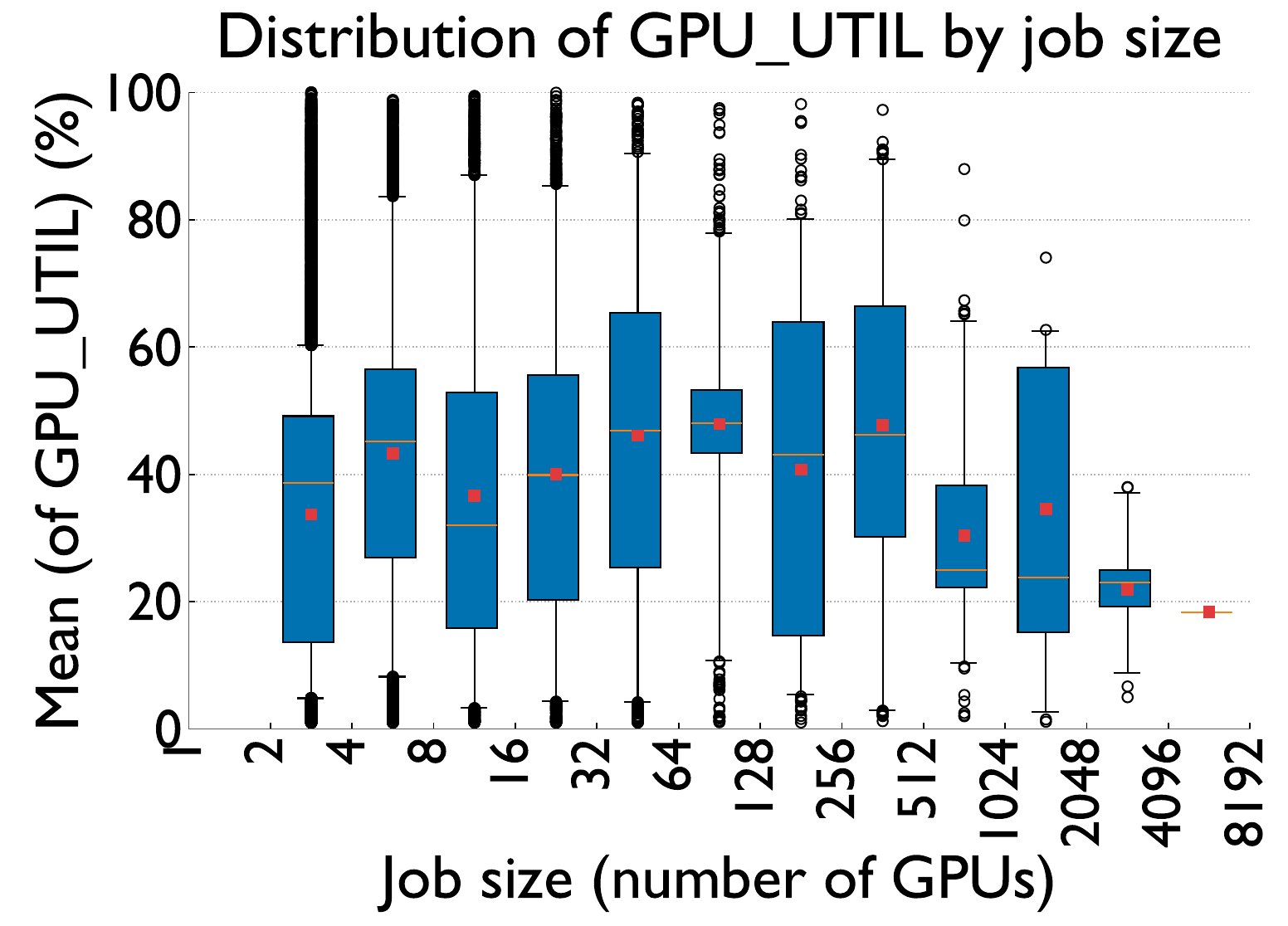}
  \caption{Left: number of jobs by job size (GPUs). A y-axis break is used to display the tall first bin while preserving readability of subsequent bins. Each bar represents a range of job sizes (e.g., the 8-16 bin includes all jobs that requested more than 8 and up to 16 GPUs). Middle: distribution of job duration (hours on log scale) by job size. Right: distribution of per-job mean GPU\_UTIL (\%) by job size. Red dots on the box plots indicate means, orange lines indicate medians, and open circles denote outliers.}
  \label{fig:num_gpus_overview}
  \vspace{-0.05in}
\end{figure*}

\subsection{Correlating job-level counter values}

We analyze correlations among job-level means of hardware counters to characterize per-job resource-usage behavior. We compute the job-level mean, $\mathit{M}(j)$, by averaging
the counter values over time, $t_j$, for each GPU in a job,
and then taking the mean across all GPUs, $g_j$, assigned to that job:
\vspace{0.2in}
\begin{equation}
  \eqnmarkbox[color_1]{m_out}{\mathit{M}(j)} =
  \eqnmarkbox[color_2]{mean_gpu}{\frac{1}{g_j}
    \sum_{g=1}^{g_j}}
  \left(
  \eqnmarkbox[color_3]{mean_time}{\frac{1}{t_j}
    \sum_{t=1}^{t_j} C_{g,t}}
  \right)
  \label{eq:mean}
\end{equation}
\annotate[yshift=.6em]{above,left}{m_out}{job-level mean\\counter value}
\annotate[yshift=-.25em]{below,left}{mean_gpu}{average over GPUs}
\annotate[yshift=.5em]{above,right}{mean_time}{average over time}
\vspace{0.2in}

where $C_{g,t} $ is the counter value for GPU $g$ at time $t$.

To quantify relationships without assuming linearity or normality, we use Spearman’s rank correlation. We compute pairwise correlation across jobs and report the full matrix. We compute all correlations on the same job set used throughout this paper.

\section{Overview of GPU workloads on Perlmutter}
\label{sec:overview}
In this section, we present an overview of our monitoring data and the
job-level characteristics of Perlmutter.
Our analysis examines job size, utilization patterns, usage of different GPU FP pipelines, HBM capacity used, and interconnect activity.
These aspects are central to understanding how workloads use system
resources.

\subsection{Overall resource usage}
Job size is an important factor that shapes workload composition.
On Perlmutter, most jobs are small, while large jobs consume a
disproportionate share of resources. Relating job size to job count, duration and GPU\_UTIL provides a baseline characterization of job behavior on the GPU partition.

The left plot in Figure~\ref{fig:num_gpus_overview} demonstrates that 64,183 jobs (85\%) allocate four GPUs (a single node). Each GPU node has four GPUs on Perlmutter. There are no jobs in the first bin (1–2 GPUs) because even when users request fewer than four GPUs, Slurm allocates all four GPUs on the node on Perlmutter. The job size in the subsequent bins decreases as job size increases.

The middle plot in Figure~\ref{fig:num_gpus_overview} demonstrates the job durations for different ranges of job sizes. The circles in the box plot represent outliers. The area within the box represents the interquartile range (IQR), defined as the range between the 5th and 95th percentiles. Whiskers above and below the box extend to the last data point within $1.5\times IQR$, and outliers are the points beyond the whiskers. This plot indicates that single-node jobs are shorter on average (\tweakedsim 55 min) compared to the multi-node jobs. Multi-node jobs have varying durations, but most multi-node size ranges have a mean value of \tweakedsim two hours. Across sizes, the IQR is roughly 1–3 hours. The absence of jobs longer than 48 hours reflects the maximum duration allowed by Perlmutter’s queue policies.

The right plot in Figure~\ref{fig:num_gpus_overview} summarizes the per-job mean GPU\_UTIL by job size, using the same box-plot conventions as the middle plot. The job count distribution skews toward small jobs (1–4 GPUs), yet means of GPU\_UTIL improve at moderate scales and peak for 33–512 GPUs (\tweakedsim 47–48\%). At very large scales ($\geq$512 GPUs), medians drop to \tweakedsim 20\% with a reduced 95th percentile, plausibly due to extended idle time from communication and synchronization in larger jobs. Small jobs (1–4 GPUs) have lower medians than 5–8 GPUs (38\% vs 45\%), because exclusive placement leaves more allocated GPUs unused in the smallest bin.

\begin{RQcallout}
    {\bf Observations }{\RQoneAnswer{}}
\end{RQcallout}

\subsection{GPU floating-point pipe activity}
Beyond overall GPU utilization, we analyze GPU FP pipe activity (FP16\_ACTV, FP32\_ACTV, FP64\_ACTV, TNSR\_ACTV) to assess how workloads drive the compute pipelines. We address the following research question:

\begin{RQcallout}
    {\bf RQ1 }{\it \RQtwo{}}
\end{RQcallout}

\begin{figure}[h]
    \centering
    \includegraphics[width=\columnwidth]{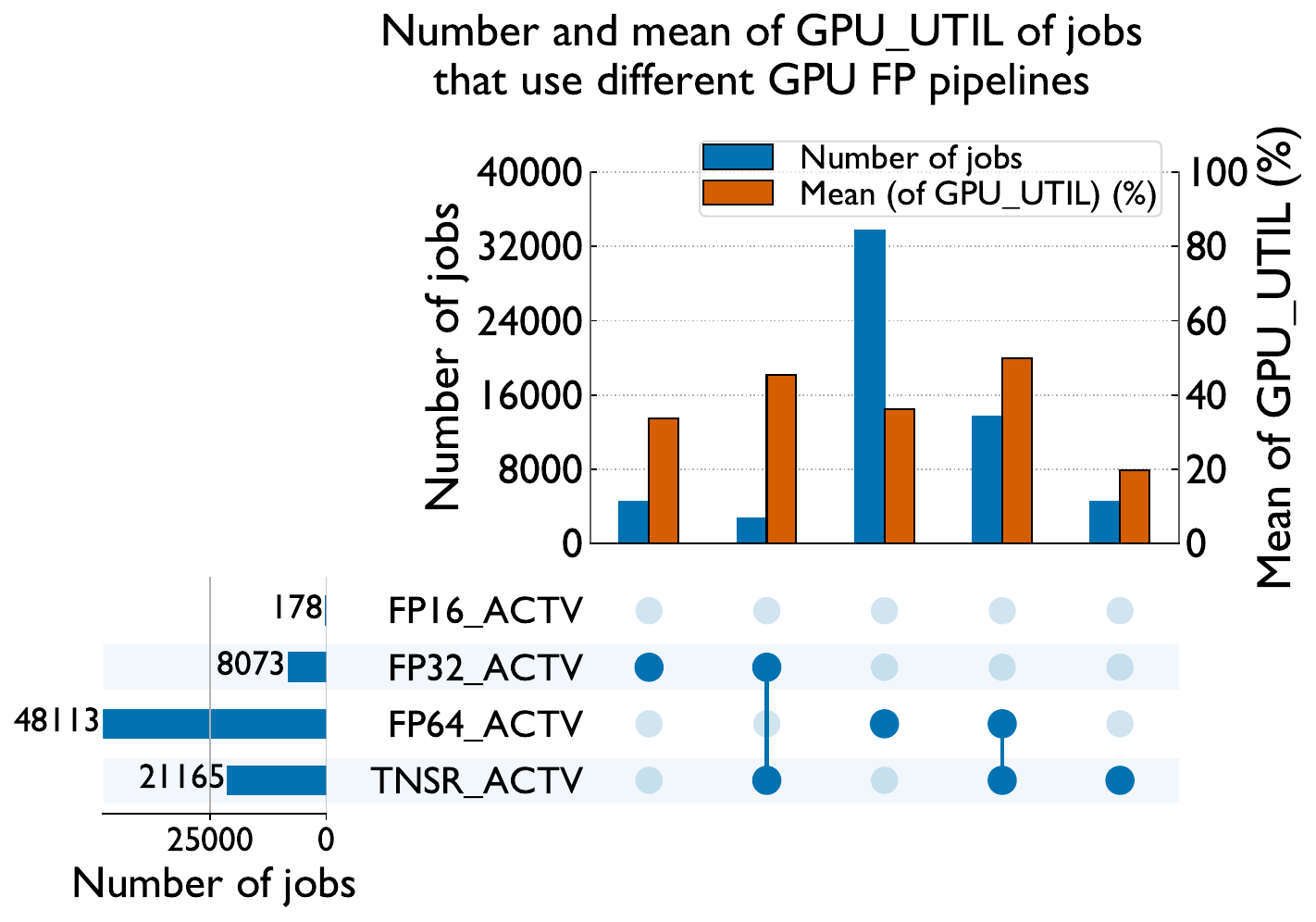}
    \caption{The number and mean GPU\_UTIL of jobs using different combinations of GPU floating-point pipes used. Each bar represents a disjoint set of jobs. Filled circles mark which FP pipes are active in that set.}
    \label{fig:precision_upsetplot}
    \vspace{-0.05in}
\end{figure}

Figure~\ref{fig:precision_upsetplot} demonstrates how much GPU jobs use different FP pipelines and how they co-occur. The top bar chart reports the number and mean GPU\_UTIL of jobs for each combination of active counters. The bottom matrix with filled circles and connecting lines denotes the counters present in each combination. Each top bar represents a disjoint set of jobs. To determine which FP pipe a job engages, we mark an FP pipe as used if the job’s mean activity for that pipe exceeds a small threshold. The threshold is the larger of the 5th percentile of per-job means for that pipe and 0.005.

FP64 dominates Perlmutter’s workload. The FP64-only intersection contains 33,675 jobs (44\% of all jobs) with a mean GPU\_UTIL of 36\%. Tensor+FP64 is the next largest intersection with 13,721 jobs (18\%) and a higher mean GPU\_UTIL of 50\%. FP32-only accounts for 4,485 jobs (6\%) at 34\% mean GPU\_UTIL. Tensor+FP32 includes 2,694 jobs (4\%) with 46\% mean GPU\_UTIL. The Tensor-only group appears with 4,463 jobs (6\%) and mean GPU\_UTIL of 20\%. Because of our small threshold, very small means of activity can fall below the threshold for some jobs and those jobs then surface as Tensor-only. Finally, FP16 is very rare on Perlmutter (178 jobs).

\begin{RQcallout}
    {\bf Answer to RQ1 }{\RQtwoAnswer{}}
\end{RQcallout}

\subsection{Peak HBM usage on 40 GB and 80 GB GPUs}
Memory (HBM) capacity is a critical constraint for GPU jobs. We analyze
peak HBM usage to assess how close jobs run to hardware limits and whether
80~GB GPUs are fully utilized. We address the following research question:

\begin{RQcallout}
    {\bf RQ2 }{\it \RQthree{}}
\end{RQcallout}

\begin{figure}[h]
    \centering
    \includegraphics[width=\columnwidth]{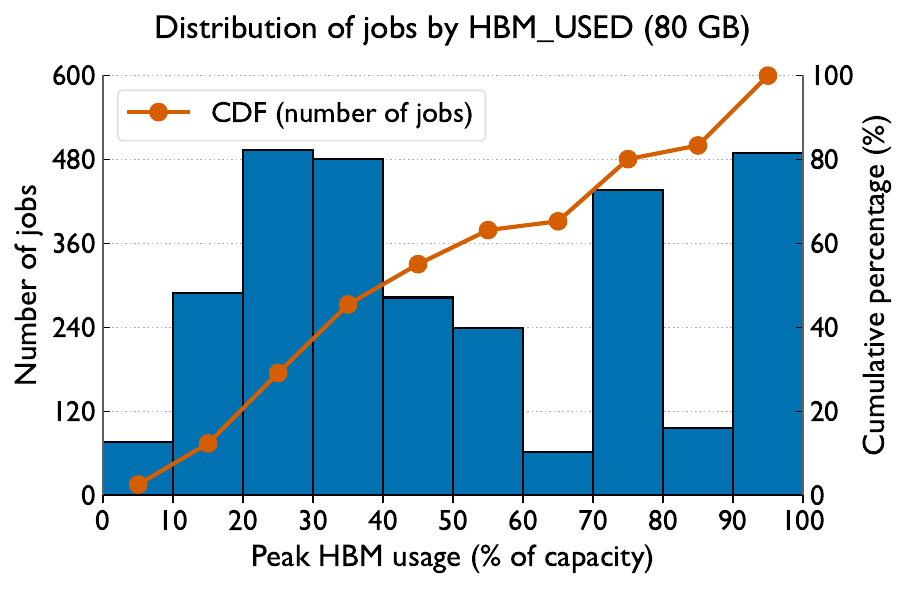}
    \caption{Distribution of peak HBM usage (normalized by capacity) for jobs that explicitly requested 80~GB GPUs. The orange line represents the CDF.}
    \label{fig:fb_used_dist}
    \vspace{-0.05in}
\end{figure}

Figure~\ref{fig:fb_used_dist} summarizes job-level peak HBM usage (HBM\_USED) for jobs that explicitly requested 80 GB GPUs. For each GPU, we compute its peak usage as the maximum HBM\_USED sample over the job’s duration divided by the GPU’s nominal capacity. The job-level peak is the maximum of these per-GPU peaks. The figure demonstrates a histogram of the resulting percentages with a CDF on the secondary y-axis.

On Perlmutter, jobs that do not request a specific memory capacity can run on either 40 GB or 80 GB GPUs. To reflect intent, we isolate the 2,947 jobs that explicitly requested 80 GB GPUs with \texttt{-C gpu\&hbm80g}. Figure~\ref{fig:fb_used_dist} reports the distribution of job-level HBM usage: 12\% of jobs use less than 20\% of capacity, 55\% use less than 50\%, 20\% use at least 80\%, and 17\% fall in the 90–100\% range. Because HBM\_USED is sampled every 10 seconds, brief spikes may be missed, therefore a safety margin should be applied when interpreting peaks.

This analysis indicates that system administrators can use historical telemetry to help users request the right amount of memory. At submission time, the system can check a user’s recent jobs. If a user requests 80 GB but prior runs on 80 GB GPUs peaked below 40 GB, the system can display an advisory prompt to consider a 40 GB GPU instead. This may reduce wait times on scarce 80 GB GPUs and improve overall throughput.

\begin{RQcallout}
    {\bf Answer to RQ2 }{\RQthreeAnswer{}}
\end{RQcallout}

\section{Characterizing job types}
\label{sec:ai}

In this section, we examine the distribution of compute- versus memory-bound
samples on Perlmutter and assess how job-level total energy consumption
differs by job type. This section addresses the following questions:

\begin{RQcallout}
    {\bf RQ3 }{\it \RQfive{}}
\end{RQcallout}
\begin{figure*}[h]
    \centering
    \par\medskip
    \vspace{-0.05in}
    \includegraphics[height=2.31in]{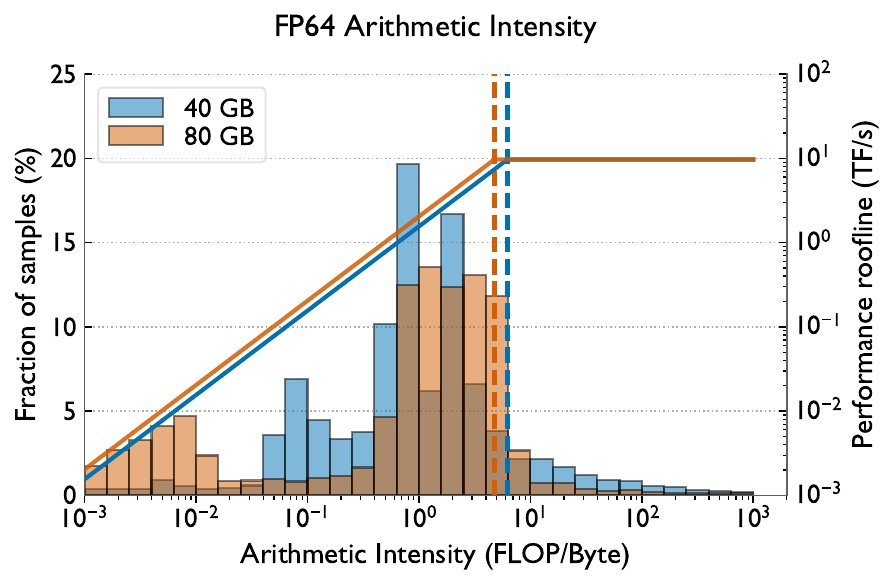}
    \includegraphics[height=2.31in]{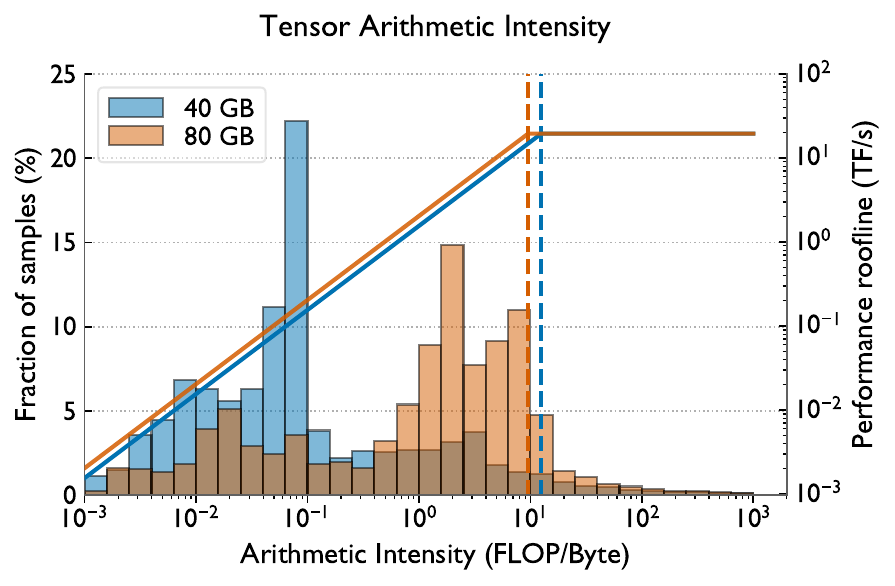}
    \caption{Distribution of arithmetic intensity by GPU floating-point pipes (40~GB versus 80~GB GPUs). Bars represent the fraction of samples for FP64 (left) and tensor (right). Vertical lines mark the capacity-specific balance points (ridges), and the right axis overlays the corresponding rooflines.}
    \label{fig:ai_40gb_80GB}
    \vspace{-0.05in}
\end{figure*}
Figure~\ref{fig:ai_40gb_80GB} presents per-sample arithmetic-intensity distributions for the FP64 and tensor pipes on NVIDIA A100 GPUs with 40 GB and 80 GB HBM. We omit the FP32 distribution for brevity because it matches FP64. Each plot reports the fraction of samples per arithmetic-intensity bin on a logarithmic axis. Balance points (ridges) and roofline ceilings are annotated. Using the FP64-specific arithmetic intensity, we classify 88\% of samples as memory-bound and 12\% as compute-bound. We assign job-level labels by the majority class among a job’s samples. When aggregated by job, 81\% of jobs are memory-bound and 19\% are compute-bound. We observe similar proportions across the other FP pipes, with ~\tweakedsim 20\% of jobs classified as compute-bound and ~\tweakedsim 80\% as memory-bound.

The separation between 40~GB and 80~GB is most visible for the tensor pipe. Tensor pipe activity is mostly associated with high compute reuse (e.g., matrix multiplication). The extra HBM capacity on 80 GB GPUs supports larger batches or models, which results in more sustained tensor flops per byte.

\begin{figure}[h]
    \centering
    \vspace{-0.05in}
    \includegraphics[width=\columnwidth]{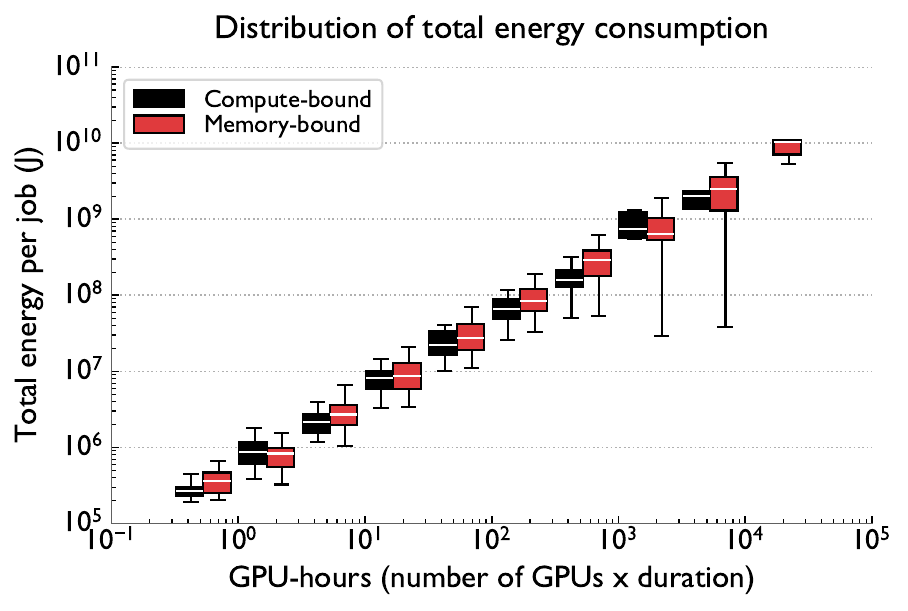}
    \caption{Distribution of total energy per job (J) versus GPU-hours with log scales on both axes. For each GPU-hour bin, side-by-side box plots compare compute-bound and memory-bound jobs classified using FP64 arithmetic intensity. GPU-hours = number of GPUs $\times$ duration of the job. White lines in boxes represent median and IQR with whiskers (outliers suppressed).}
    \label{fig:eng_temp_comp_mem}
    \vspace{-0.05in}
\end{figure}

\begin{figure*}[t]
    \centering
    \vspace{-0.05in}
    \includegraphics[height=1.47in]{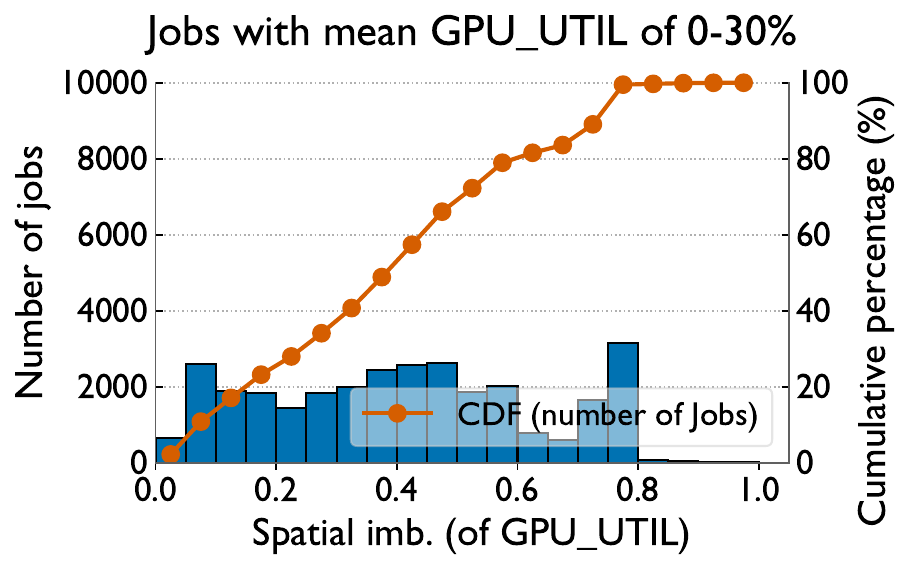}
    \includegraphics[height=1.47in]{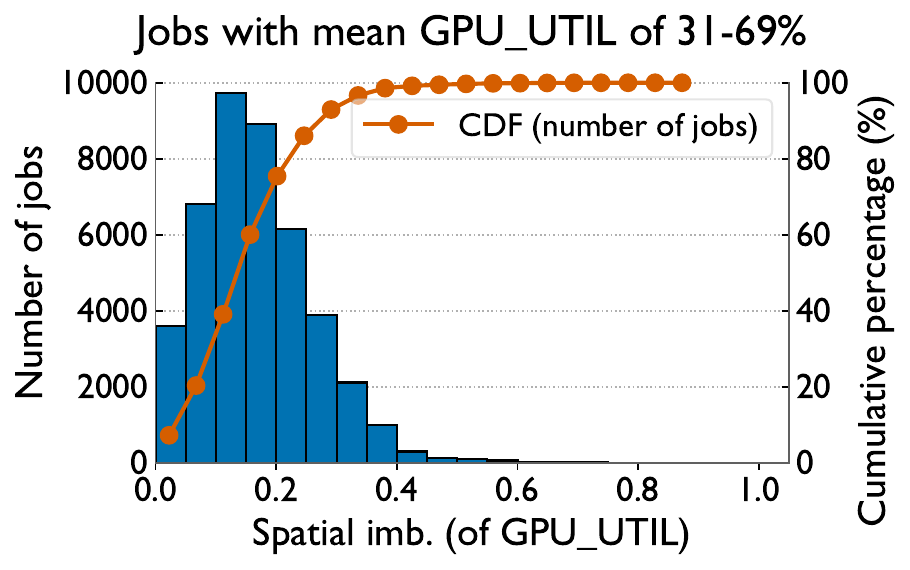}
    \includegraphics[height=1.47in]{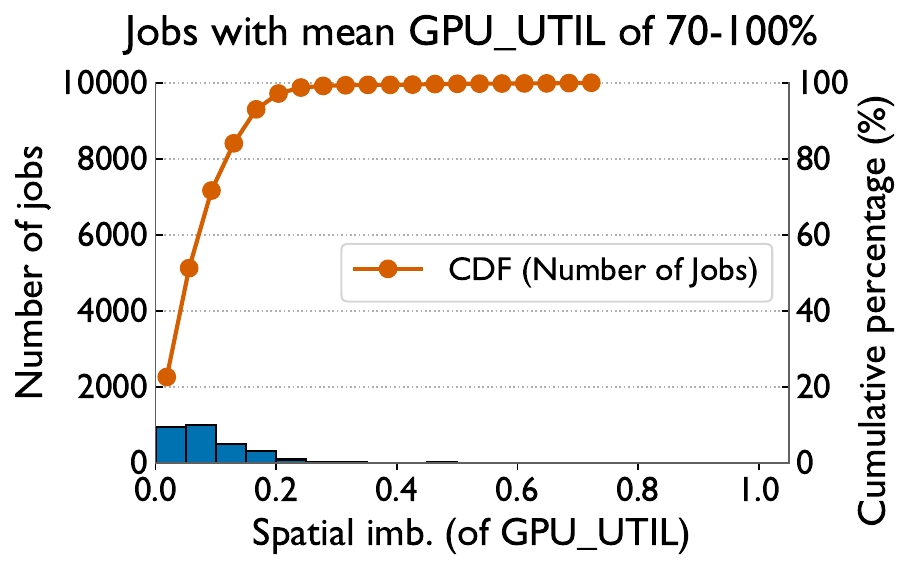}
    \caption{The plots demonstrate the distributions of spatial imbalance of GPU\_UTIL for jobs grouped by mean GPU\_UTIL (0–30\%, 31–69\%, 70–100\%)}

    \label{fig:gputil_spatial}
    \vspace{-0.05in}
\end{figure*}

We next compare total energy consumption of compute- and memory-bound jobs (classified by FP64 arithmetic intensity). In Figure~\ref{fig:eng_temp_comp_mem}, each box plot demonstrates the distribution of total energy per job (J) within a GPU-hours bin. For every bin on the x-axis, we place two box plots side by side: one for compute-bound jobs and one for memory-bound jobs so that we can compare two classes at the same GPU-hours scale. Here, we define GPU-hours as the number of GPUs allocated multiplied by the job's runtime in hours.

Two patterns emerge. First, as GPU-hours increase, the entire energy distribution shifts upward and broadens as expected from longer and larger runs. Second, at comparable GPU-hours, memory-bound jobs tend to consume more total energy than compute-bound jobs at a given GPU-hour scale (their boxes sit higher in the same bins). Antepara et al. empirically observe that off-chip HBM traffic carries a much higher energy cost than arithmetic, which can translate into higher average power for memory-bound workloads and aligns with our observation of higher total energy at comparable GPU-hours~\cite{antepara2025benchmarkdriven}.

\begin{RQcallout}
    {\bf Answer to RQ3 }{\RQfiveAnswer{}}
\end{RQcallout}

\section{Analyzing the spatial behavior of jobs}
\label{sec:spatial}
This section analyzes spatial imbalance in GPU jobs. We use a fixed one-minute window to define the metric, after comparing one, five, 15, and 30 minutes. With 15–30 minute windows, a brief spike on a single GPU can dominate the entire interval because the metric compares every sample in that interval to the most active GPU. This tends to overstate imbalance in bursty, low-utilization jobs. Shorter windows (1–5 minutes) attribute high imbalance only to the periods when it actually occurs. We therefore use a 1-minute window. We group jobs by mean GPU\_UTIL into low (\textless30\%), medium (30–70\%), and high (\textgreater70\%) categories, and we also examine the effects of FP pipeline usage and job type. This section addresses the following questions:
\begin{RQcallout}
    {\bf RQ4 }{\it \RQsix{}}
\end{RQcallout}

\begin{figure*}[t]
    \centering
    \vspace{-0.05in}
    \includegraphics[height=1.47in]{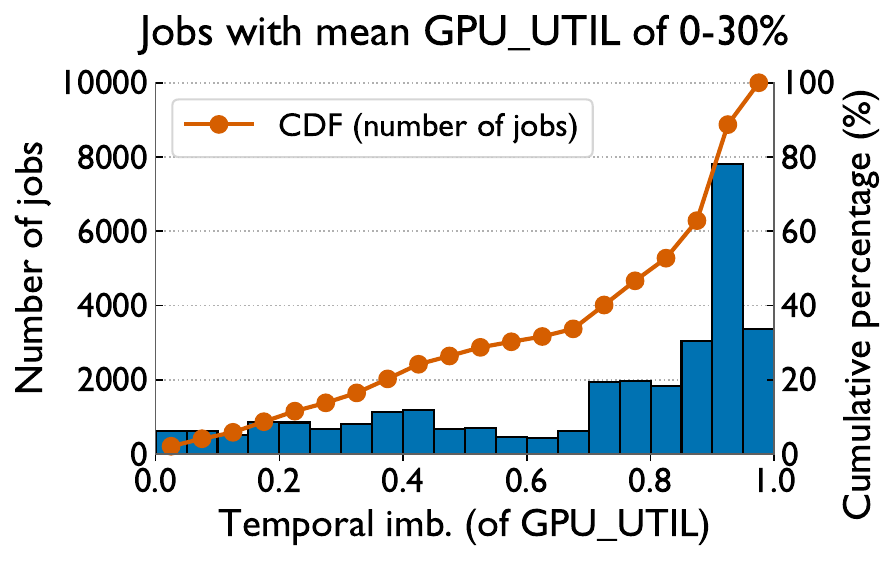}
    \includegraphics[height=1.47in]{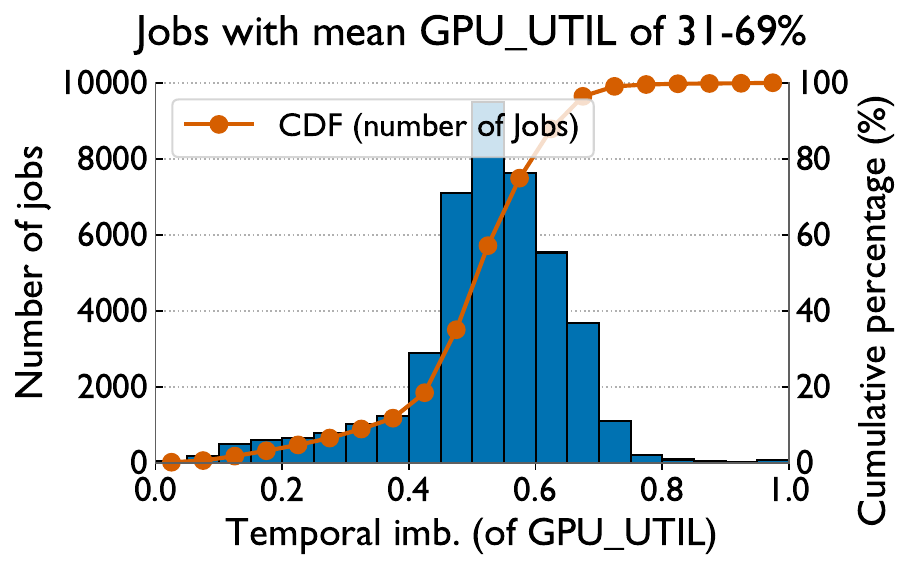}
    \includegraphics[height=1.47in]{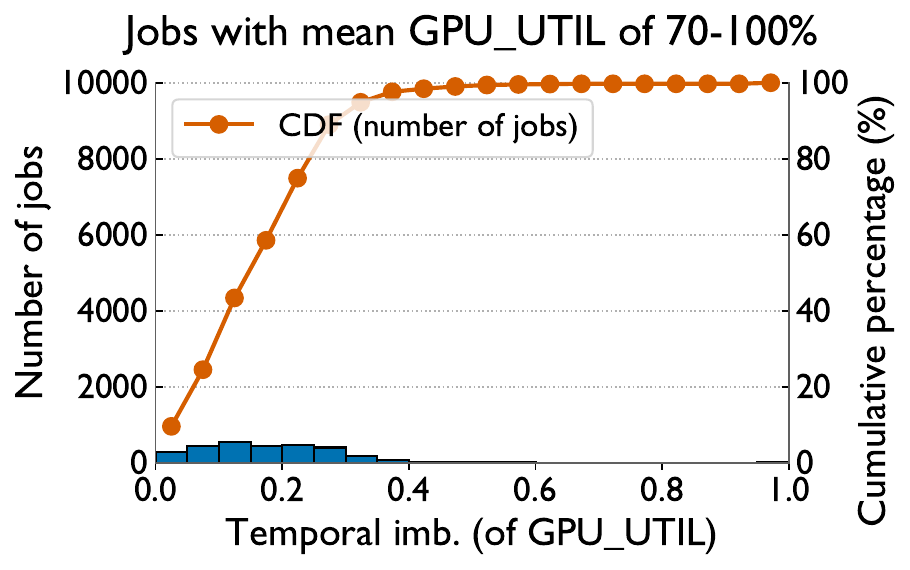}
    \caption{Distributions of temporal imbalance of GPU\_UTIL for jobs grouped by
        mean GPU\_UTIL (0–30\%, 31–69\%, 70–100\%). Low-utilization jobs concentrate at
        high imbalance, while high-utilization jobs are tightly clustered at low imbalance.}
    \label{fig:gputil_temporal}
    \vspace{-0.05in}
\end{figure*}

Figure~\ref{fig:gputil_spatial} demonstrates the spatial imbalance distribution of
GPU\_UTIL across three utilization ranges. Low-utilization jobs
(left) exhibit the highest imbalance (peaking at 0.78),
which indicates a significant unevenness in resource distribution. We observe that
only 34\% of jobs in this category fall below 0.30, and 16.5\% are at or above 0.70. Medium-utilization jobs (middle)  are more
balanced (86\% fall below 0.30). High-utilization jobs (right) have less spatial imbalance
and most of them concentrate in the lower range: a total of 97.1\% fall below
0.30.

Additionally, we observe that 12\% of all 4-GPU jobs never used three of their GPUs (max of GPU\_UTIL = 0). This pattern is consistent with Perlmutter’s exclusive GPU-node allocation, where the jobs allocate all four GPUs even if they use only one of them.

\begin{figure}[h]
    \centering
    \par\medskip
    \vspace{-0.05in}
    \includegraphics[width=\columnwidth]{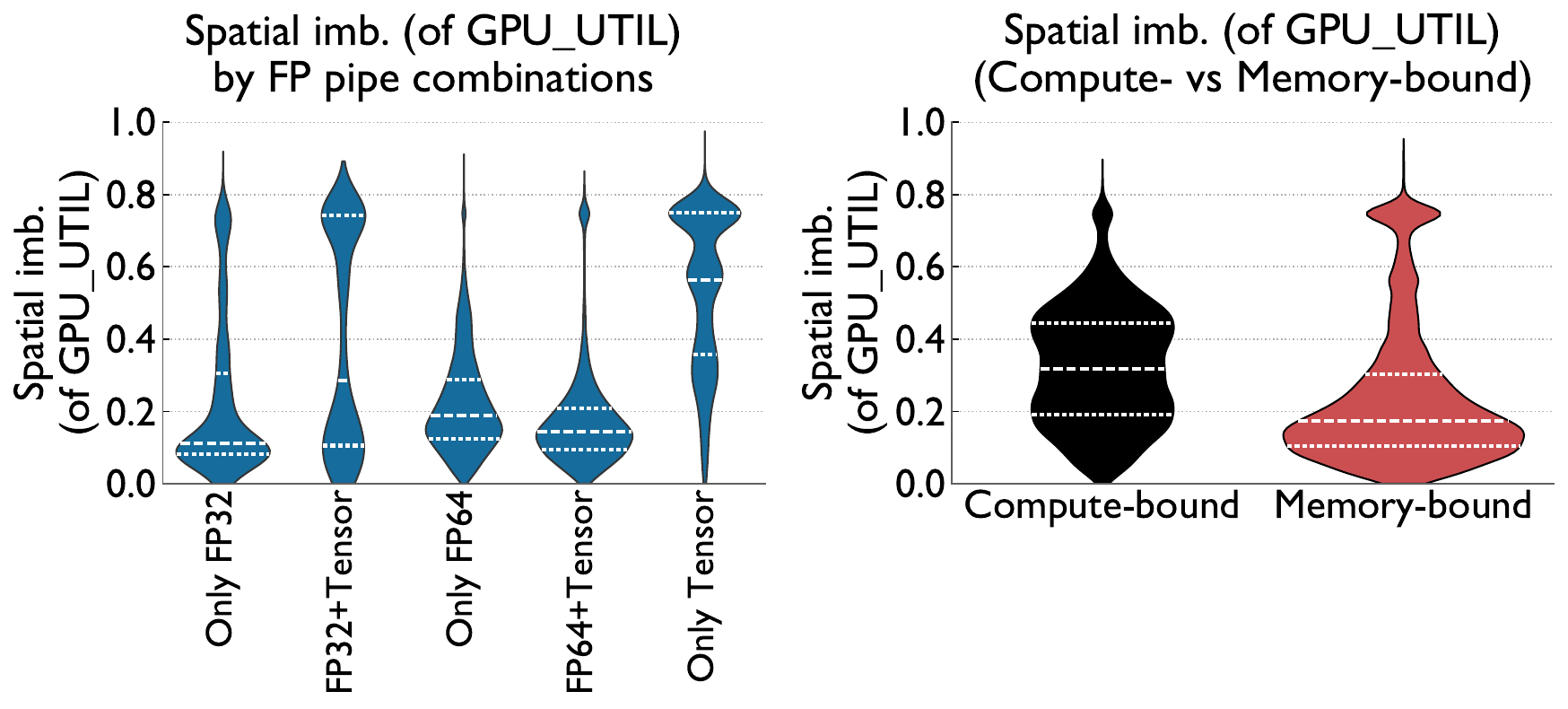}
    \caption{The plots demonstrate spatial imbalance by FP pipes used and by job type. Left: spatial imbalance of GPU\_UTIL grouped by FP pipes used. Right: spatial imbalance of GPU\_UTIL by job type. Each violin illustrates the job-level distribution and is area-normalized (width encodes density); internal lines mark quartiles.}
    \label{fig:SI_pf_roof}
    \vspace{-0.05in}
\end{figure}

Figure~\ref{fig:SI_pf_roof} relates spatial imbalance of GPU\_UTIL to FP pipeline usage (left plot) and job type (right plot). We normalize the density
of each violin so that the total area remains consistent across all
violins. White lines mark quartiles. The left plot reports spatial imbalance of GPU\_UTIL for exclusive FP pipe group derived from Figure~\ref{fig:precision_upsetplot}. The right plot reports spatial imbalance of GPU\_UTIL by job type.

Figure~\ref{fig:SI_pf_roof} demonstrates that Tensor-only jobs exhibit the greatest spatial imbalance (median 0.56, IQR 0.36–0.75; p95 0.75). FP32+Tensor is also elevated (median 0.28, IQR 0.10–0.74). By contrast, FP32-only (median 0.11) and FP64-dominated combinations are lower: FP64+Tensor (median 0.14, IQR 0.09–0.20) and FP64-only (median 0.18, IQR 0.12–0.28).

Using FP64-derived arithmetic intensity, compute-bound jobs exhibit higher typical imbalance than memory-bound jobs (median values of 0.31 and 0.17, respectively). Overall, tensor-heavy groups and compute-bound workloads typically have higher spatial imbalance, whereas memory-bound jobs tend to be lower on average.

\section{Analyzing the temporal behavior of jobs}
\label{sec:temporal}
In this section, we analyze the distribution of temporal imbalance of GPU\_UTIL and quantify its dependence on FP pipes used and job types. Temporal imbalance captures how
consistently a job sustains GPU activity over its runtime. This section
addresses the following question:

\begin{RQcallout}
    {\bf RQ5 }{\it \RQseven{}}
\end{RQcallout}

Figure~\ref{fig:gputil_temporal} demonstrates the distribution of temporal imbalance
of GPU\_UTIL for jobs grouped by mean GPU\_UTIL: low
(\textless30\%), medium (31\%-69\%), and high (\textgreater70\%).
Low-utilization jobs (left) are broadly
spread, with most clustered at high imbalance: only 13.7\% fall below a
temporal imbalance value of 0.30, while 66.3\% exceed 0.70. Medium-utilization jobs (middle) are more balanced. Their distribution shifts lower and tightens around 0.5, with 6.5\% below 0.30 and 3.6\% above 0.70. Finally, high-utilization jobs (right) concentrate at low imbalance and 88.9\% of high-utilization jobs fall below temporal imbalance value of 0.30, and only 0.3\% exceed 0.70.

\begin{figure}[h]
    \centering
    \par\medskip
    \vspace{-0.05in}
    \includegraphics[width=\columnwidth]{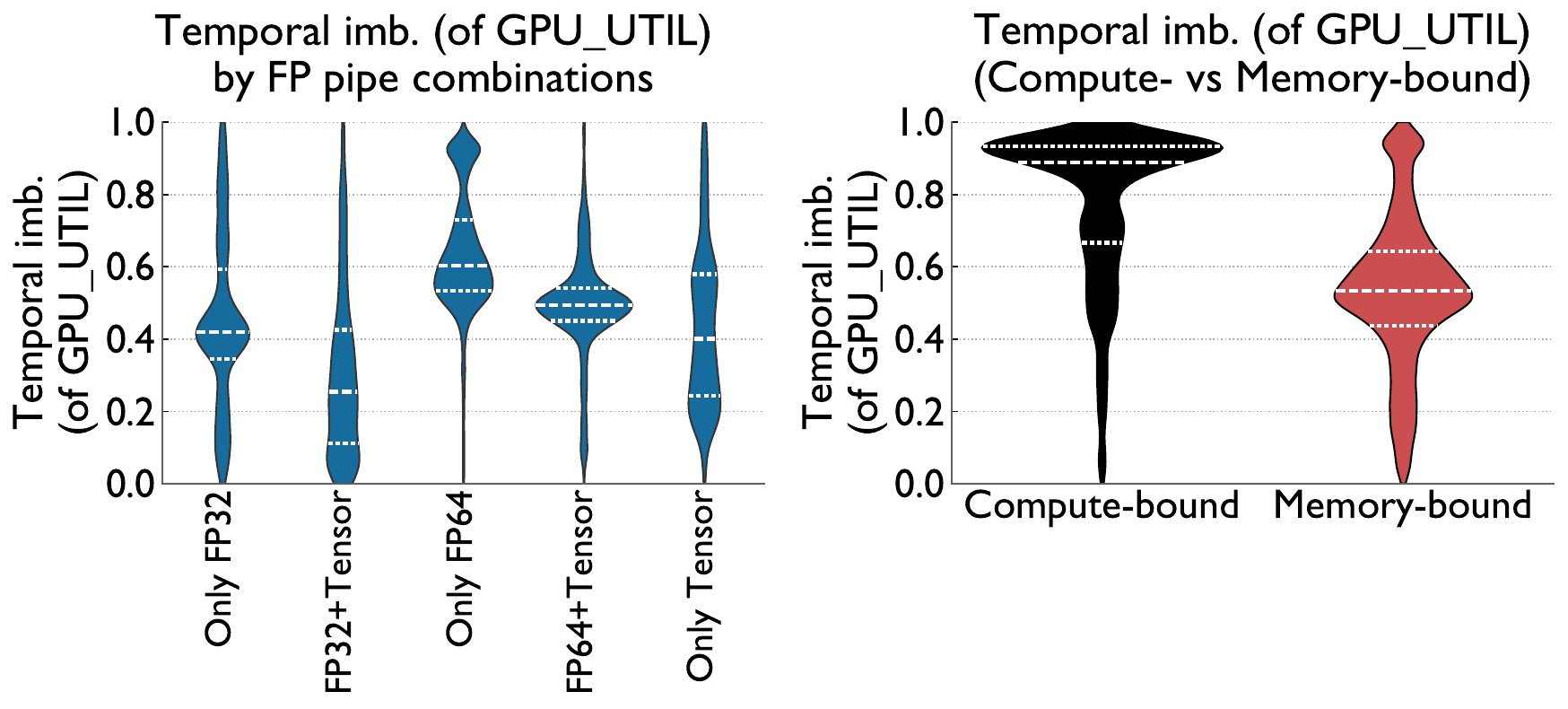}
    \caption{The plots demonstrate temporal imbalance by FP pipes used and by job type. Left: temporal imbalance of GPU\_UTIL grouped by FP pipes used. Right: temporal imbalance of GPU\_UTIL by job type. Each violin illustrates the job-level distribution and is area-normalized (width encodes density); internal lines mark quartiles.}
    \label{fig:roofline_TI}
    \vspace{-0.05in}
\end{figure}

Figure~\ref{fig:roofline_TI} demonstrates temporal imbalance distributions by FP pipe used (left plot) and job type (right plot). Each violin is an equal-area density with quartiles, as in Figure~\ref{fig:SI_pf_roof}.
Temporal imbalance varies with both FP pipes (left plot) and job type (right plot). FP64-only exhibits the highest and relatively tight temporal imbalance (median 0.60, IQR 0.53–0.73). FP64+Tensor (median 0.50, IQR 0.45–0.54), FP32-only (median 0.42, IQR 0.34–0.60) and Tensor-only (median 0.40, IQR 0.24–0.58) are moderate, while FP32+Tensor is lowest (median 0.25, IQR 0.11–0.43). By job type, compute-bound jobs exhibit markedly higher temporal imbalance than memory-bound jobs (medians 0.89 vs. 0.53; IQRs 0.67–0.93 vs. 0.44–0.64).

\begin{RQcallout}
    {\bf Answer to RQ5 }{\RQsevenAnswer{}}
\end{RQcallout}

\section{Analyzing relationships between counters}
\label{sec:relationship}
In this section, we examine how mean GPU\_UTIL changes with FP64\_ACTV and
DRAM\_ACTV for compute- versus memory-bound jobs (via job-level heatmaps), and we
compute a Spearman correlation matrix over job-level means to summarize how all
counters relate to one another. This section addresses the
following research questions:
\begin{RQcallout}
    {\bf RQ6 }{\it \RQeight{}}
\end{RQcallout}

\subsection{GPU utilization of compute- versus memory-bound jobs}

We measure how job-level mean GPU\_UTIL changes with compute and memory activity. For compute, we use FP64\_ACTV because it is the most commonly used pipeline. For memory, we use DRAM\_ACTV, which tracks the achieved bandwidth on the A100 at a fixed HBM clock.

\begin{figure}[h]
    \centering
    \includegraphics[width=0.49\columnwidth]{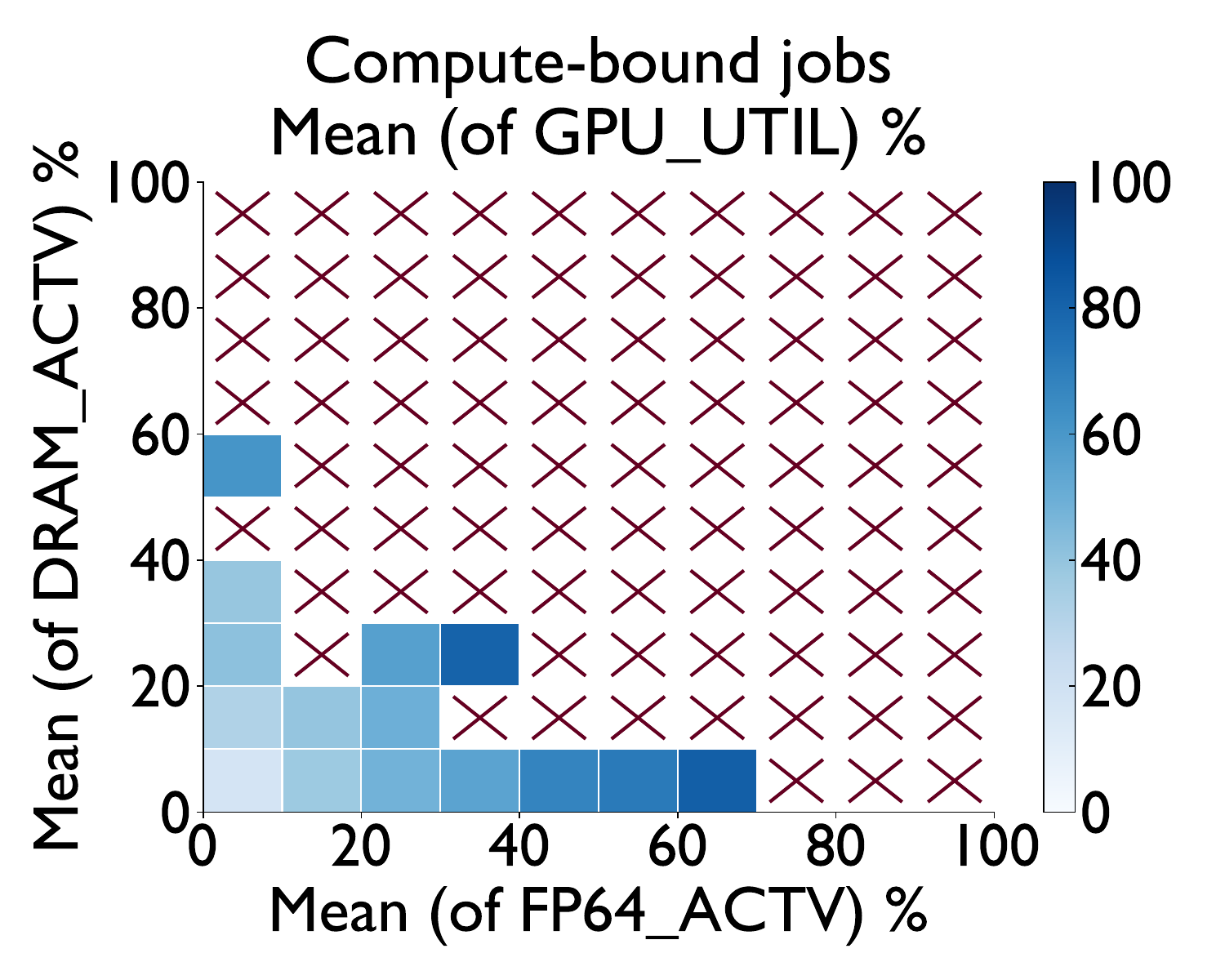}
    \includegraphics[width=0.49\columnwidth]{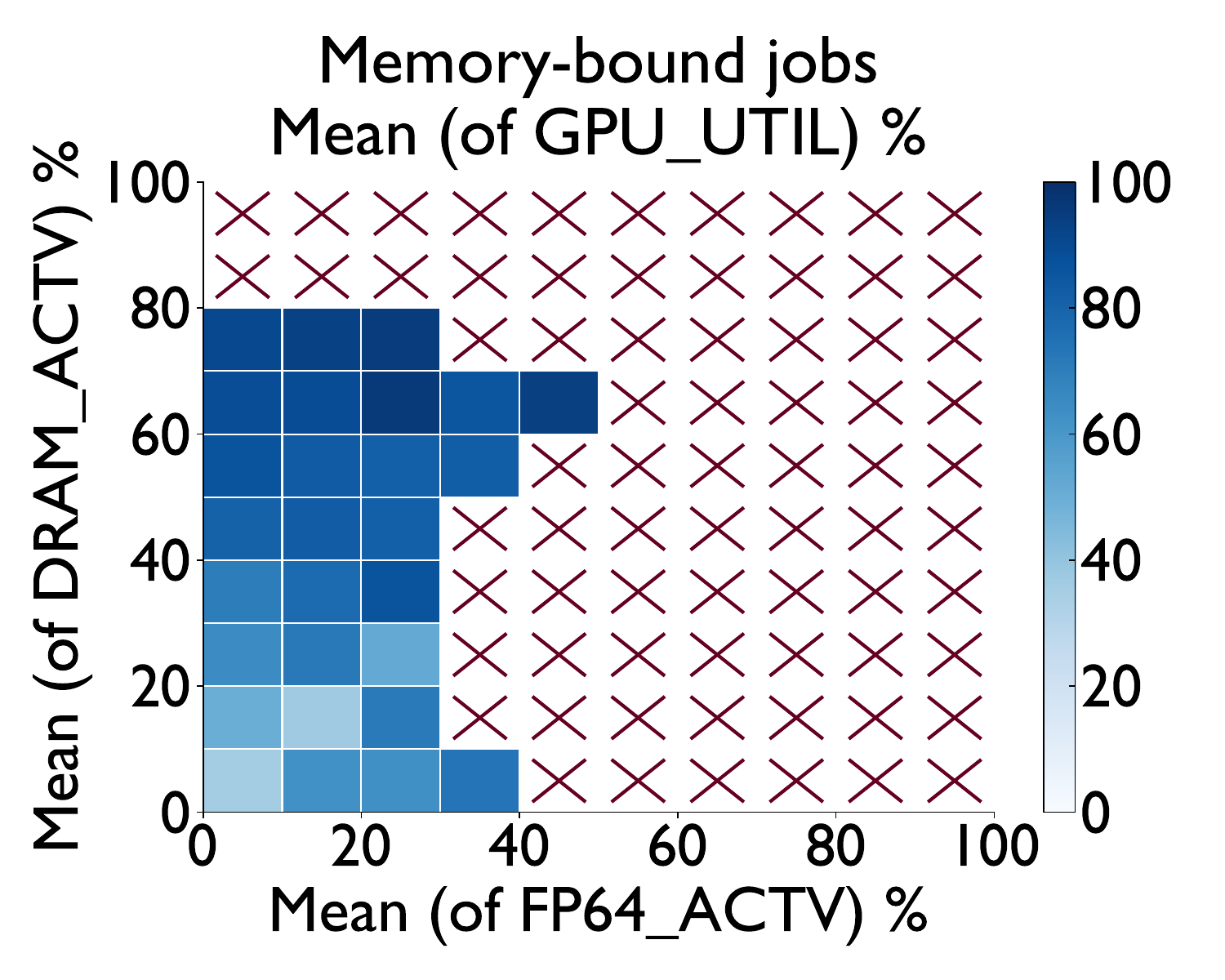}
    \caption{The plots demonstrate the job-level heatmaps of mean GPU\_UTIL as a function of FP64\_ACTV and DRAM\_ACTV. We bin jobs by their mean FP64\_ACTV (x-axis) and mean DRAM\_ACTV (y-axis); color encodes the mean GPU\_UTIL of jobs in each bin. Left: compute-bound jobs. Right: memory-bound jobs. Crosses indicate empty bins.}
    \label{fig:dram_pseudo64_gputil}
\end{figure}

Figure~\ref{fig:dram_pseudo64_gputil} plots the resulting heatmaps. For compute-bound jobs (left), mean GPU\_UTIL increases primarily with FP64\_ACTV across a wide range of DRAM\_ACTV, with corresponding correlations of 0.508 for FP64\_ACTV and 0.102 for DRAM\_ACTV. In memory-bound jobs (right), mean GPU\_UTIL increases with both axes. Bins with DRAM\_ACTV \textgreater 40\% sustain higher GPU utilization even when mean FP64\_ACTV is low. This pattern aligns with job-level correlations of mean GPU\_UTIL with FP64\_ACTV (0.513) and DRAM\_ACTV (0.570).

These patterns are consistent with roofline reasoning. The performance of the compute-bound jobs (to the right of the ridge) is limited by peak flop/s. Therefore, incremental changes in delivered HBM bandwidth have little leverage. In the memory-bound regime (left of the ridge), sustained throughput improves either by delivering more bandwidth (higher DRAM\_ACTV) or by increasing arithmetic intensity (more useful flops per byte), which typically co-moves with FP pipe activity. We emphasize that these are associations, not causal proofs. The exact magnitudes depend on factors such as kernel mix and cache behavior.

\subsection{Correlation among hardware counters}
To understand how hardware counter values co-vary at the job level, we compute a Spearman correlation matrix over job-level means of the counters (Eq.~\ref{eq:mean}). We order  the heatmap by the absolute correlation with GPU\_UTIL.

\begin{figure}[h]
    \centering
    \includegraphics[width=1\columnwidth]{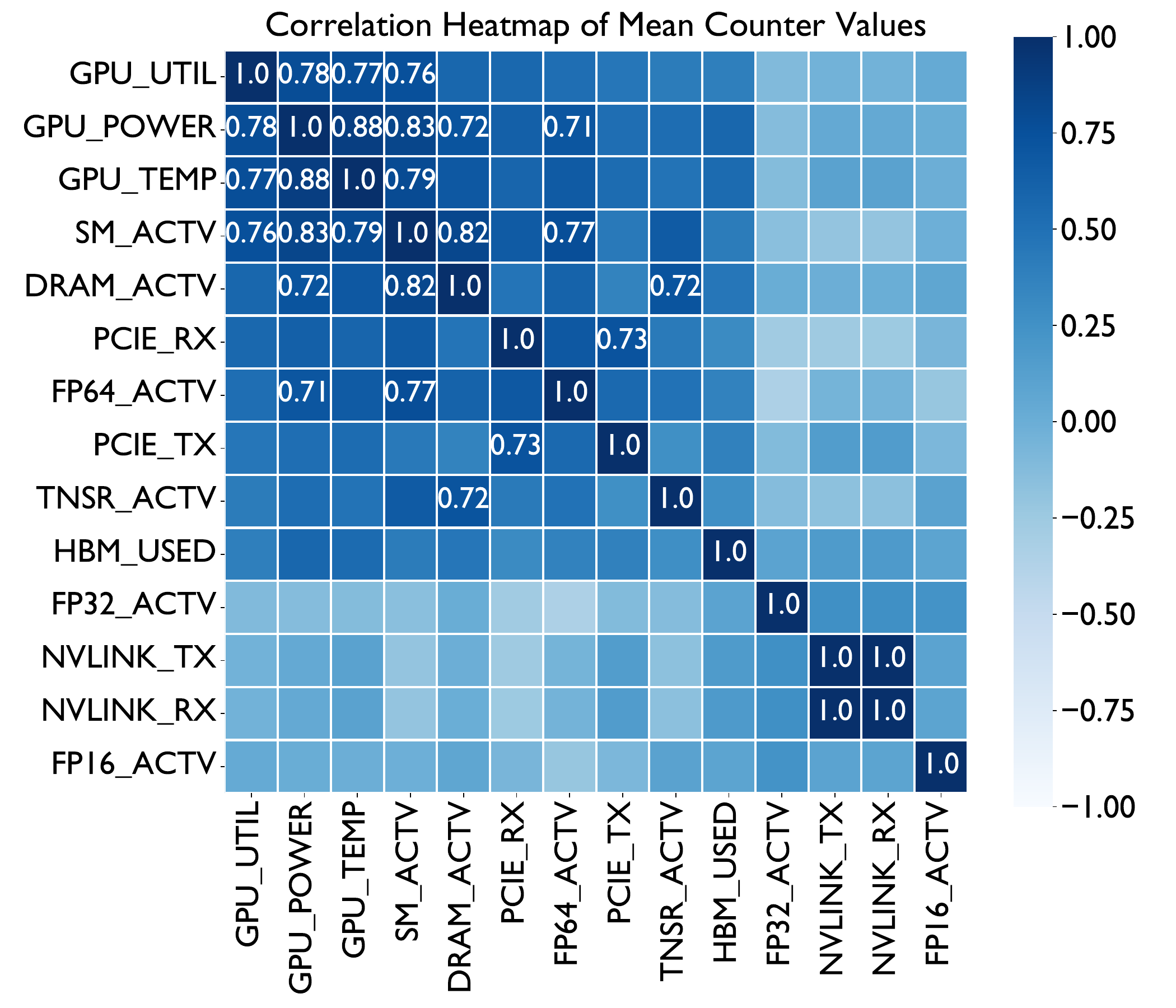}
    \caption{The plot presents a Spearman correlation heatmap of job-level mean counters. We order cells by absolute correlation with GPU\_UTIL. Only strong correlations ($|\rho|\ge 0.7$) are annotated.}
    \label{fig:correlation_heatmap_all}
\end{figure}

Figure~\ref{fig:correlation_heatmap_all} demonstrates that mean GPU\_UTIL correlates most strongly with
GPU\_POWER (0.78), and GPU\_TEMP (0.77), and SM\_ACTV (0.76). We calculate GPU\_POWER by dividing the total energy consumption per job (TOTAL\_ENG) by GPU count and job duration. These counters co-vary strongly across jobs: higher active compute is accompanied by higher power draw and device temperature. DRAM\_ACTV correlates positively as well (0.58), consistent with memory activity supporting sustained utilization.

SM\_ACTV correlates strongly with DRAM\_ACTV (stalled warps are counted as active) and with FP64\_ACTV, while having a weak negative association with FP32\_ACTV (\textminus0.11). This negative sign reflects workload composition: FP64-dominant and FP32-dominant jobs are largely disjoint groups, so raising one pipe’s activity tends to lower the other at the job-level mean, yielding a small inverse relationship.

NVLINK\_TX and NVLINK\_RX are nearly symmetric (0.99), consistent with bidirectional data exchange in intra-node collectives. In contrast, PCIE\_RX and PCIE\_TX are less symmetric (0.73) and only moderately related to GPU\_UTIL (0.57 and 0.46), likely because PCIe traffic is often directional (e.g., inter-node communication, host-to-device staging, checkpoints).

\begin{RQcallout}
    {\bf Answer to RQ6 }{\RQeightAnswer{}}
\end{RQcallout}

\section{Discussion}
\label{sec:discussion}
In this section, we summarize implications from a system-wide analysis of resource usage on Perlmutter GPU nodes using a variety of DCGM counters.

\vspace{0.06in}
\noindent\textbf{Leveraging tensor cores.}
Tensor pipe activity is typically associated with higher overall GPU utilization. While this relationship is associative rather than causal, it motivates adopting or upgrading tensor-enabled libraries/kernels where appropriate. For FP64 workloads (the most common on Perlmutter), targeted mixed-precision can improve device activity.

\vspace{0.06in}
\noindent\textbf{Job type determines the effective optimization.}
Our roofline-based analysis indicates that compute-bound jobs align primarily with FP pipe activity, whereas memory-bound jobs benefit from increases in both DRAM activity and FP pipe activity that raises arithmetic intensity. Practically, for compute-bound jobs, optimization should prioritize kernel efficiency, instruction-level parallelism, and increased floating-point issue. For memory-bound jobs, optimization should prioritize memory throughput and reuse (tiling, caching, reduced precision where valid) to shift arithmetic intensity rightward.

\vspace{0.06in}
\noindent\textbf{Energy as a scheduling signal.}
At comparable GPU-hour scales, memory-bound jobs tend to consume more total energy than compute-bound jobs. For operators, this suggests energy-aware placement and capping policies when memory-bound workloads dominate a node. For users, reducing memory traffic (and increasing reuse) can yield both performance and energy benefits.

\vspace{0.06in}
\noindent\textbf{Imbalance diagnostics for users and operators.}
Spatial and temporal imbalance concentrate in low-utilization jobs and in specific FP pipe mixes. Exposing spatial/temporal dashboards in user-facing job monitors can help identify load imbalance (rebalance partitions, adjust batch sizes, or request fewer GPUs), while operators can use this information to guide co-scheduling.

\vspace{0.06in}
\noindent\textbf{Properly sizing allocations for HBM requests.}
Peak HBM usage is typically below device capacity for many jobs that explicitly request 80-GB GPUs. A lightweight submission-time advisory could inspect a user’s recent per-job peak HBM usage and recommend an appropriate memory tier (e.g., 40-GB vs 80-GB) before the scheduler queues the job. Such guidance may reduce wait times on scarce 80-GB nodes and improve overall throughput, without enforcing any policy.

\vspace{0.06in}
\noindent\textbf{Limitations.}
Our results summarize one month of Perlmutter DCGM data sampled every 10 seconds and rely on job-level aggregation. The sampling interval is set by system administrators and cannot be changed by users, so brief spikes may be missed. Our analysis is intentionally limited to what can be inferred from production monitoring without code-level profiling, so function call paths and thread-level behavior are not captured. While the methodology is broadly applicable, comparable job-level telemetry is hard to obtain across sites due to access permissions and monitoring differences. Higher-frequency traces, cross-site validation, and complementary code-level profiling are valuable directions for future work.

\section{Related work}
\label{sec:related}
Many monitoring tools have been developed to monitor performance of HPC
systems such as LDMS~\cite{ldms}, ParMon~\cite{parmon}, and
SuperMon~\cite{supermon}.  These tools typically continuously
monitor the compute nodes and collect performance data about the jobs running
on them.  They can provide information about CPU, GPU, memory, I/O usage and
power, energy consumption. The collected data are typically
time-series-based.
Perlmutter utilizes the LDMS for system
monitoring and can provide GPU-level time-series data for every job
running on the compute nodes.

Several studies have leveraged monitoring data for different purposes. Some
studies use monitoring data to predict the runtime of
jobs~\cite{7776517, aaziz2018modeling, costello:arxiv2020}, resource
usage~\cite{emeras2017evalix, sirbu2016power, gimenez:sc2017}, and performance variability~\cite{nichols:ipdps2022, bhatele:ipdps2020}.  Others apply monitoring
data for  application fingerprinting~\cite{li2023arcode, ates2018taxonomist}.
These studies indicate that monitoring data can be used to identify and cluster jobs running the same
application by analyzing the monitoring data. Additionally, many studies
utilize monitoring data to identify performance anomalies. Many researchers
apply various ML techniques for automatic anomaly
detection~\cite{proctor, ruad, ozer2020characterizing}.  These studies
highlight various ways monitoring data can provide insights into
HPC system behavior ~\cite{netti2021conceptual, netti2022operational}.

This paper focuses on resource usage analysis and workload characterization.
Many studies have characterized the performance and resource utilization of HPC
and cloud systems using monitoring data. Most studies have primarily
focused on the CPU and memory utilization to characterize
workloads~\cite{amvrosiadis2018diversity, cortez2017resource,
    shen2015statistical, chen2010analysis, mishra2010towards,
    reiss2012heterogeneity, ren2012workload, di2012characterization, leon:sc2016, bhatele:ijhpca2010}. For example,
Peng et al. analyze memory utilization on two HPC systems and investigate
spatial and temporal imbalances of jobs~\cite{peng2021holistic}.  Similarly, multiple papers analyze temporal and spatial characteristics of jobs
by using counters such as CPU utilization, GPU utilization, and memory usage
in Perlmutter~\cite{li2023analyzing, sencanldms}.

Additionally,
several studies have focused on I/O and storage
utilization~\cite{shen2015statistical, patel2019revisiting, ren2012workload}.
For example, Patel et al. analyze I/O behaviors in large-scale storage systems~\cite{patel2019revisiting}.  Several studies explore power and
energy consumption~\cite{shen2015statistical, amvrosiadis2018diversity,
    li2023analyzing, ilager2023data, chu2024generic}.  Chu et al. compare the
energy profiles of Machine Learning (ML) and traditional workloads and find
that the ML jobs have higher power usage and failure
rates~\cite{chu2024generic}.  Ilager et al.~\cite{ilager2023data} analyze
energy and temperature in cloud data centers.  Hu et al.~\cite{hu2021characterization} characterize DL
workloads on multiple data centers and find that single-GPU
jobs have a weaker impact on cluster usage compared to multi-GPU
jobs.

While previous works have highlighted inefficiencies such as underutilized CPU or
GPU resources and memory imbalances, they often lack detailed GPU-specific
analyses because they use a limited variety of GPU-related performance
counters. Additionally, most prior work uses node-level data, whereas we analyze
GPU-level data in this study.  Our study provides a more fine-grained analysis
of GPU hardware counters with improved metrics to analyze the behavior of jobs.

\section{Conclusion}
\label{sec:conclusion}
This study offers a job-level, GPU-centric characterization of workload behavior using previously underutilized DCGM counters. Spatial and temporal imbalance expose uneven and unsteady use of allocated GPUs; the roofline-based job characterization clarifies differences in energy consumption; FP pipeline analysis differentiates job behavior; and counter correlations indicate which signals move together. Collectively, these results support practical choices about optimization targets and scheduling policies on Perlmutter, and the methodology is applicable to other GPU-accelerated systems.

\section*{Acknowledgment}
This material is based upon work supported in part by the National Science Foundation
under Grant No.~2047120.  This research used resources of the National Energy
Research Scientific Computing Center (NERSC), a U.S.~Department of Energy (DOE)
Office of Science User Facility, operated under Contract No.~DE-AC02-05CH11231
using NERSC award DDR-ERCAP0034262.

\bibliographystyle{IEEEtran}
\bibliography{./bib/cite,./bib/pssg}

\appendices
\def\IPDPSADINCLUDED{1}
\appendixAD

\section{Overview of Contributions and Artifacts}
In this section we list the contributions this work makes.

\subsection{Paper's Main Contributions}

\begin{description}
    \item[$C_1$] We analyze 75,703 Perlmutter jobs over one month using largely underexplored DCGM counters and a robust job-attribution pipeline.
    \item[$C_2$] We label jobs as compute-bound or memory-bound using a roofline criterion.
    \item[$C_3$] We propose a time-windowed spatial-imbalance metric to capture uneven usage across a job's GPUs.
    \item[$C_4$] We classify jobs by FP pipeline activity and quantify how utilization and spatial/temporal imbalance vary by class.
    \item[$C_5$] Using a broad set of system-wide DCGM counters, we quantify job-level relationships and identify which counters co-vary with overall GPU utilization.
\end{description}

\subsection{Computational Artifacts}

\begin{description}
    \item[$A_1$] Scripts and Jupyter notebooks used to perform the analyses and figures in the paper (one notebook per experiment). Publicly available at https://github.com/hpcgroup/ldms-analysis-scripts-ipdps26
    \item[$A_2$] Monitoring data collected from the Perlmutter supercomputer. This dataset is restricted because it was collected at a U.S. Department of Energy site, so it cannot be shared publicly.
\end{description}

\begin{center}
    \begin{tabular}{rll}
        \toprule
        Artifact ID & Contributions       & Related        \\
                    & Supported           & Paper Elements \\
        \midrule
        $A_1$       & $C_1$, $C_2$, $C_3$ & Figures 1-11   \\
                    & $C_4$, $C_5$        &                \\
        \midrule
        $A_2$       & $C_1$, $C_2$, $C_3$ & Table 1        \\
                    & $C_4$, $C_5$        & Figures 1-11   \\
        \bottomrule
    \end{tabular}
\end{center}

\section{Artifact Identification}

\newartifact
$A_1$ is the analysis code artifact. It consists of 11 Jupyter notebooks (one per figure) that reproduce Figures~1--11 from the paper using the Perlmutter telemetry dataset ($A_2$).

\artrel
$A_1$ implements the analysis workflow used in the paper, including aggregation of DCGM telemetry to the job level, roofline-based compute/memory labeling, spatial and temporal imbalance metrics, FP-pipeline-based job classification, and system-wide relationship analyses across counters. When executed with $A_2$ as input, $A_1$ reproduces the plots and statistics that support contributions $C_1$--$C_5$ (Figures~1--11).

\artexp
Running the notebooks regenerates the paper figures. Each notebook filename is prefixed with its corresponding figure number:
\begin{itemize}
    \item Fig.~1: 1\_overall\_resource\_usage.ipynb
    \item Fig.~2: 2\_gpu\_fp\_pipe\_activity.ipynb
    \item Fig.~3: 3\_peak\_hbm\_usage.ipynb
    \item Fig.~4: 4\_arithmetic\_intensity.ipynb
    \item Fig.~5: 5\_total\_energy\_consumption.ipynb
    \item Fig.~6: 6\_dist\_spatial\_imb.ipynb
    \item Fig.~7: 7\_spatial\_imb\_fp\_compute\_memory.ipynb
    \item Fig.~8: 8\_dist\_temporal\_imb.ipynb
    \item Fig.~9: 9\_temporal\_imb\_fp\_compute\_memory.ipynb
    \item Fig.~10: 10\_heatmap\_fp64\_dram\_gputil.ipynb
    \item Fig.~11: 11\_correlation\_heatmap.ipynb
\end{itemize}

\arttime
Assuming $A_2$ is already collected, it takes 15--30 minutes to create the software environment and configure dataset paths, 40--45 hours total to execute all 11 notebooks (approximately four hours per notebook), 15--30 minutes to verify that generated figures match the paper.

\artin
\artinpart{Hardware}
A CPU compute node on Perlmutter is required.

\artinpart{Software}
Python with Jupyter is required. The notebooks rely on standard data and plotting packages (e.g., NumPy, pandas, pyarrow, matplotlib).

\artinpart{Datasets / Inputs}
The notebooks consume the restricted Perlmutter telemetry dataset ($A_2$).

\artinpart{Installation and Deployment}
Follow the repository instructions to create the Python environment (e.g., via pip), then execute the notebooks.

\artcomp
Retrieve the DCGM and Slurm data and configure the dataset paths. Next, set up the Python/Jupyter environment according to the repository specifications. Then, execute the notebooks to generate Figures~1--11. Finally, collect the generated figure files and compare them against the versions in the paper.

\artout
The primary outputs are the regenerated Figures~1--11. All paper plots and derived results are produced by running $A_1$ with $A_2$ as input.

\newartifact
$A_2$ is the curated telemetry dataset that is utilized for the system-wide study in this paper. It is a data-only artifact and does not execute experiments by itself.

\artrel
$A_2$ contains NVIDIA DCGM counter time series collected across Perlmutter GPU nodes, along with associated job metadata. These data are consumed by the analysis scripts in $A_1$ and therefore support contributions $C_1$--$C_5$.

\artexp

This artifact provides the input data needed to perform analyses in the paper. The dataset contains telemetry collected on Perlmutter over the study period (March 1 to April 1, 2025).

\arttime

Assuming access is already granted on the DOE site used in the paper (Perlmutter), retrieving the dataset typically takes up to three days.

\artin

\artinpart{Hardware}
Obtaining this dataset requires authorized access to the Perlmutter supercomputer environment.

\artinpart{Software}
The telemetry is collected and curated exclusively by NERSC staff using NVIDIA DCGM (i.e., users cannot run the system-wide collection themselves). Authorized users can only retrieve the staff-provided dataset.

\artinpart{Datasets / Inputs}
This artifact is the dataset itself.

\artinpart{Installation and Deployment}
No installation is required.

\artcomp
First, NERSC staff (not users) collects the system-wide telemetry and makes it available to authorized users. Next, the dataset is copied to the execution location (e.g., a project filesystem or scratch). Finally, the dataset is validated by checking the time range, schema, and basic counts before running any analyses. The primary experimental parameters represented in this dataset are the system (Perlmutter) and the collection window.

\artout
The output of this artifact is the curated telemetry dataset itself (restricted), including the telemetry tables and accompanying metadata.
All plots, tables, and derived results in the paper are produced by running $A_1$ using $A_2$ as input.

\end{document}